\documentstyle[preprint,aps,version2]{revtex}
\tightenlines

\newcommand\semidirect{\mathbin{\hbox{\hskip2pt\vrule height 5.0pt depth -.3pt
width .25pt \hskip-2pt$\times$}}}

\font\zfont = cmss10 
\newcommand\ZZ{\hbox{\zfont Z\kern-.4emZ}}

\newcommand{\bib}[5]{#1 \it #2\bf#3\ \rm(19#5)\ #4}
\newcommand{\PLB}{Phys. Lett. \bf B}
\newcommand{\PRD}{Phys. Rev. \bf D}
\newcommand{\NPB}{Nucl. Phys. \bf B}
\newcommand{\IJMPA}{Int. J. Mod. Phys. \bf A}

\newcommand{\be}{\begin{equation}}
\newcommand{\ee}{\end{equation}}
\newcommand{\beq}{\begin{eqalignno}}
\newcommand{\eeq}{\end{eqalignno}}
\newcommand{\bea}{\begin{eqnarray}}
\newcommand{\eea}{\end{eqnarray}}
\newcommand{\br}{\overline}
\newcommand{\del}{\partial}
\newcommand{\eqr}[1]{(\ref{#1})}
\newcommand{\rarr}{\rightarrow}
\newcommand{\al}{\alpha}
\newcommand{\bt}{\beta}
\newcommand{\bata}{scaling coefficient}
\newcommand{\batas}{\bata s}
\newcommand{\djn}{N=$\hbox{2\kern-.45em /}$ }
\newcommand{\djns}{N=$\hbox{2\kern-.45em /}^*$ }
\newcommand{\gm}{\gamma}
\newcommand{\lam}{\lambda}
\newcommand{\Lam}{\Lambda}

\newcommand{\eps}{\epsilon}
\newcommand{\hlf}{{1\over2}}
\newcommand{\ot}[1]{\tilde{#1}}

\newcommand{\nf}{{N_f}}
\newcommand{\nc}{{N_c}}

\newcommand{\fq}{\fqn{r}}
\newcommand{\aq}{\aqn{s}}
\newcommand{\Miti}{$M_{s}^{\; r}$}
\newcommand{\Mitim}{M_{s}^{\; r}}

\newcommand{\fQn}[1]{Q^{#1}}
\newcommand{\aQn}[1]{\ot{Q}_{#1}}
\newcommand{\fqn}[1]{q_{#1}}
\newcommand{\aqn}[1]{\ot{q}^{#1}}

\newcommand{\Lgr}{{\cal L}}
\newcommand{\Op}{{\cal O}}
\begin{document}


\pagestyle{empty}
\preprint{RU--95--2}
\medskip
\preprint{hep-th/9503121}
\medskip
\preprint{March 1995}

\vspace{0.3in}

\begin{title}
Exactly Marginal Operators and Duality\\
in \\
Four Dimensional N=1 Supersymmetric Gauge Theory
\end{title}

\medskip

\author{Robert G. Leigh and Matthew J. Strassler
\thanks{Work supported in part by the Department of Energy, contract
DE--FG05--90ER40559.}}

\begin{instit}
Department of Physics and Astronomy\\
Rutgers University, Piscataway, New Jersey 08855
\end{instit}


\begin{abstract}

We show that manifolds of fixed points, which are generated by exactly
marginal operators, are common in N=1 supersymmetric gauge theory.  We
present a unified and simple prescription for identifying these
operators, using tools similar to those employed in two-dimensional
N=2 supersymmetry. In particular we rely on the work of Shifman and
Vainshtein relating the $\bt$-function of the gauge coupling to the
anomalous dimensions of the matter fields.   Finite N=1 models, which
have marginal operators at zero coupling, are easily identified using
our approach.  The method can also be employed to find manifolds of
fixed points which do not include the free theory; these are seen in
certain models with product gauge groups and in many non-renormalizable
effective theories. For a number of our models, S-duality may have
interesting implications.  Using the fact that relevant perturbations
often cause one manifold of fixed points to flow to another, we propose
a specific mechanism through which the N=1 duality discovered by Seiberg
could be associated with the duality of finite N=2 models.

\bigskip\bigskip\bigskip\bigskip

\centerline{(Submitted to Nuclear Physics B)}

\end{abstract}


\newpage
\pagestyle{plain}

\narrowtext

\section{Introduction}
\label{sec:intro}

The study of conformal field theory in two dimensions has an extensive
history.  Among the features common in these theories are extended
manifolds of fixed points, often called fixed lines, fixed planes, {\it etc.}
These manifolds are generated by ``exactly marginal'' operators;
at a fixed point with an exactly marginal operator $\Op$, the
addition of the operator to the Lagrangian $\delta\Lgr=h\Op$
leads to a new fixed point for a continuous range of $h$.
To prove $\Op$ is exactly marginal is generally very difficult;
marginality at $h=0$ is insufficient, since
the dimension of $\Op$ may vary with $h$.
In two dimensions one has large classes of soluble models where the
dimensions of operators are exactly known and fixed lines can be proven
to exist.  However, there are also models with N=2 supersymmetry,
which, although not always soluble, have enough symmetry
that it is possible to prove that an operator is exactly
marginal\cite{lance,chiralrings}. As we will show, this is also true
in four-dimensional N=1 supersymmetry.

The existence of marginal operators
in N=1 supersymmetry is implicit in the work of various authors
in the context of perturbatively finite
models \cite{finiteearly}--\cite{finiteSUfive}.  (For finite models,
Refs.~\cite{finitelate,PandS} reach much the same
conclusions as we do but use more complex techniques to arrive at them;
the approach of Piguet and Sibold and collaborators\cite{PandS}
is related to ours, though they use a very different language.)
An example has also been identified in an interacting
theory \cite{NSKItwo}.
We will present a unified and simplified description of these phenomena,
and will display many new examples, demonstrating that
marginal operators are to be found throughout supersymmetric gauge theory.
Many of these generate manifolds of fixed points with properties
worthy of further investigation.  In particular we will see several
models in which the phenomenon of strong-weak coupling duality may be
studied, and we will see a suggestive relation between N=2 duality
and the N=1 duality studied by Seiberg \cite{NAD}.

Our method, which we will explain more thoroughly below, can be
easily summarized. We use certain properties of $d=4$, N=1 supersymmetric
gauge theories which are similar to those of $d=2$, N=2 supersymmetric
theories\cite{chiralrings}.  Holomorphy of the
superpotential implies severe restrictions; in particular, couplings
of chiral fields in the superpotential are not perturbatively
renormalized\cite{nonrenthm,Holo}.  Non-perturbative renormalizations
of the superpotential are restricted by holomorphy
\cite{Holo,Seibone,ILS}. Still, any physical coupling {\it is} renormalized,
and its running can
be expressed in terms of its canonical dimension and the anomalous
dimensions of the fields that it couples.  That is, corresponding to the
superpotential $W=h\phi_1\dots\phi_n$ there is a $\bt$-function
\be\label{betah}
\bt_h \equiv \frac{\del h(\mu)}{\del \ln\mu}
= h(\mu)\Big( -d_W+\sum_k \Big[d(\phi_k)+\hlf\gm(\phi_k)\Big]\Big)
\equiv h(\mu)A_h
\ee
where $d_W$ is the canonical dimension of the superpotential, $d(\phi_k)$
is the canonical dimension of the field $\phi_k$ and $\gm(\phi_k)$
is its anomalous mass dimension.  We will refer to $A_h$ as
a \bata; it is related to the physical dimension of the
operator $\phi_1\dots\phi_n$.  The Wilsonian gauge couplings undergo
renormalization only at the one-loop level; the physical running
gauge coupling then has an exact $\bt$-function\cite{SV}
\be\label{shifvain}
\bt_g \equiv \frac{\del g(\mu)}{\del \ln\mu} =
- f(g[\mu])\bigg(\Big[3C_2(G)- \sum_k T(R_k)\Big]
                   +\sum_k T(R_k) \gm(\phi_k) \bigg)\equiv f(g[\mu])A_g
\ee
where $C_2(G)$ is the quadratic Casimir of the adjoint representation,
$T(R_k)$ is the quadratic Casimir of the representation in
which $\phi_k$ appears, and $f(g)$ is a function of the gauge coupling
which may have a pole at large $g$ but is otherwise smooth and positive.
The derivation of these statements (which we will review below)
requires only the scale dependence of the Wilsonian effective action
and the chiral (Konishi) anomaly.

We now examine the conditions for a fixed point.  If we have $n$
independent couplings $g_i$ in the theory, we have $n$ $\bt$-functions
$\bt_i(g_1,\dots,g_n)$.  At a fixed point, the
$\bt$-functions must all vanish.  This
requirement imposes $n$ conditions on the $n$ couplings.  If these
conditions are independent, we expect at most isolated solutions.
However, it may happen that
some of the $\bt$-functions are linearly dependent.  If $p<n$ is
the number of linearly independent $\bt$-functions, then
$\bt_i=0$ imposes only $p$ conditions on the $n$ couplings, and
we expect the generic solution to be an $n-p$ dimensional submanifold
in the space of couplings.  Of course, in no case are we guaranteed that
any solutions can be found; we can only say that if a solution exists,
the solution space will generically have dimension $n-p$.  Translation
within an extended space of fixed points corresponds to varying a
marginal coupling constant.  Each such coupling constant is associated
with a marginal operator in the theory which
remains exactly marginal within the manifold of fixed points.

As an example of the power of these observations, we review a typical
two-dimensional Landau-Ginsburg model \cite{chiralrings}.
The model has $n$ identical chiral
superfields $\phi_k$ and a superpotential $W=\lam\sum_k (\phi_k)^n$.
We may take $\lam$ to be real and positive by redefining the fields.
The symmetries require that all fields have the same anomalous dimension
$\gm$.  Using \eqr{betah}, $d_W=1$ and $d_k=0$, one finds
a non-trivial $\bt$-function $\bt_\lam\propto -1+\hlf n\gm(\lam)$.  If a
superconformal fixed point exists, then we have
$\gm(\lam_*)=\frac{2}{n}$. For this fixed point to be
stable there must exist an $\eps>0$ such that for $0<\delta<\eps$,
$\gm(\lam_*-\delta)<\frac{2}{n}<\gm(\lam_*+\delta)$.

Now let us add the operator $\delta W = h\phi_1\phi_2\dots\phi_n$
to the theory.  This operator is marginal at the conjectured fixed
point, and it preserves the flavor symmetry, so
the anomalous dimensions of the fields remain equal to one another.  The
coupling $h$ has $\bt$-function
$\bt_h\propto -1+\hlf\sum_k\gm(\lam,h) \propto \bt_\lam$,
so from the two $\bt$-functions we have
only one condition on the couplings $\lam,h$.  We therefore expect
a complex curve of fixed points, specified by $\gm(\lam,h)=\frac{2}{n}$,
which  passes through the point $(\lam,h)=\lam_*,0$ but extends
out into the $(\lam,h)$ plane.  This conclusion is unavoidable as
long as $\gm(\lam,h)$ is a continuous function,
as can be seen from the stability condition for the fixed point
at $h=0$.

We now turn to four-dimensional supersymmetric gauge theory.
Using \eqr{betah} and \eqr{shifvain} we have searched for models
which have coupling constants whose $\bt$-functions are linearly dependent.
We will present a number of models in Sec.~\ref{sec:finite},
many of which have appeared
in the literature, which have manifolds of fixed points
intersecting the point where the gauge coupling
vanishes; such theories can be analyzed at weak coupling.
Many (if not all) of these N=1 models, including both vector-like and chiral
theories, are finite, at least in the sense in which N=4 supersymmetric
theories are finite. That is, the anomalous dimensions of all chiral
operators vanish, so that the effective action has no ultraviolet
divergences; there can still be divergent non-chiral operator
renormalizations. There is an extensive literature on finite
N=1 quantum field theories in four dimensions
\cite{finiteearly}--\cite{finiteSUfive}, and
detailed lists of these models are given in \cite{listfin}.

We then turn to manifolds of non-perturbative fixed points.
We have found a class of theories that contains both
weakly and strongly coupled models.  The study of this
set of examples (Sec.~\ref{sec:weaktostrong}) gives us confidence to apply
our techniques outside the perturbative regime.  We then
display a number of other cases in Sec.~\ref{sec:nonrenorm}, including
models with chiral matter content, which have the potential for fixed
manifolds at strong coupling.  In this situation the candidate marginal
operator is perturbatively non-renormalizable; consequently our approach
is on unstable ground in these models and our conclusions are
conjectural.   Still, we will  argue that our methods
apply to effective field theories with a supersymmetric cutoff,
subject to certain conditions, so that renormalizability is not
a requirement.

Even if our methods do apply, they do not rule out the possibility
in most interacting models that there are no fixed points anywhere.
However, certain $SU(N)$ supersymmetric gauge theories with matter in the
fundamental representation were recently studied by
Seiberg \cite{NAD}, who conjectured the existence of a wide class of
interacting fixed points using a variety of arguments.
(The existence of interacting fixed points for large gauge groups
can be verified perturbatively.)  The case of $SO(N)$ and $Sp(N)$ gauge
groups was outlined in \cite{NAD} and the former
has been fully explored in \cite{NSKItwo}.  As we will show, in
certain cases these fixed points have exactly marginal operators,
and so we are reasonably confident that fixed lines exist in these models.
The fixed points studied by Seiberg also have
the special property \cite{NAD} that they have at least two
descriptions involving different gauge groups with different
representation content.
This ``N=1 duality'' is useful to us in certain cases where
it gives additional evidence that the formulas \eqr{betah} and
\eqr{shifvain} are non-perturbatively valid.

Finally we will address the issue of renormalization group flow
from one manifold of fixed points to another.  We will present
a wide variety of examples in Sec.~\ref{sec:RGflow}.  We observe that
it is possible that
dualities present in certain theories are partially preserved under the
renormalization group flow.  In particular we will present evidence
in Sec.~\ref{sec:duality} for a connection between duality
of N=2 finite models and the N=1 duality observed in \cite{NAD}.

\section{Vanishing of Scaling Anomalies and Anomalous Dimensions}
\label{sec:SVderiv}

The conditions  \eqr{betah} and \eqr{shifvain}
follow from simple considerations about scaling invariance in
four-dimensional supersymmetric gauge theory.  This was first
shown by Shifman and Vainshtein \cite{SV}; we now review their
argument, slightly adapted for our purposes.  We will follow
its presentation with a discussion of its significance and limitations.

\subsection{Conditions for fixed points and marginal operators}
\label{subsec:derive}
Consider a non-Abelian gauge theory with $n$ chiral fields $\phi_i$
and a superpotential $W(\phi_i)$.
Classically the theory is scale invariant except for terms in the
superpotential; the derivative of the supercurrent multiplet (which
includes the energy-momentum tensor, the supercurrent, and the chiral
$R$ current \cite{Scurrents})  gives
\be
\br D^{\dot \al}J_{\al\dot \al}\Big|_{{\rm classical}}
= \frac{1}{3}D_\al\left(3W-\sum_{i=1}^n
\phi_i\frac{\del W}{\del \phi_i}\right)
\ee
Quantum mechanically there are additional terms from the scaling
anomalies \cite{Sanomalies}
associated with the gauge one-loop $\bt$-function and
the anomalous dimensions of the matter fields.  These anomalies
cause additional dependence of the supercurrent operator on the Wilsonian
cutoff $\mu$.
\be\label{DJanom}
\br D^{\dot \al}J_{\al\dot \al}=
\br D^{\dot \al}J_{\al\dot \al}\Big|_{{\rm classical}}
- \frac{1}{3}D_\al\left( \frac{b_0}{32\pi^2} W_\bt W^\bt
           + \br \frac{D^2}{8} \sum_{i=1}^n
		    \gm_i Z_i \phi^{\dag}_i e^V \phi_i\right)
\ee
where $W_\bt$ is the gauge field strength superfield,
$b_0=3C_2(G)- \sum_i T(R_i)$ with notation as in \eqr{shifvain},
$Z_i(\mu)$ is the wave function renormalization  of the field
$\phi_i$ and
$\gm_i = -\del\ln Z_i/\del\ln\mu$ is its anomalous mass dimension.
We have assumed
here that the fields $\phi_i$ do not mix under renormalization.  If
they do mix, then a field redefinition should be performed so that the
matrix of anomalous dimensions is diagonal, following which this
derivation will apply.

The equation of motion for each field $\phi_i$, multiplied by $\phi_i$
and corrected to account for the chiral (Konishi) gauge anomalies
\cite{Sanomalies,Konanom} of the theory, is
\be\label{EOM}
\frac{\br{D}^2}{4} Z_i \phi^{\dag}_i e^V \phi_i =
\frac{1}{16\pi^2} T(R_i) W_\bt W^{\bt}
 + \phi_i \frac{\del W}{\del \phi_i} \ .
\ee
Substituting this into the supercurrent anomaly, we find
\be
\br D^{\dot \al}J_{\al\dot \al}=
\frac{1}{3}D_\al
\left[ -\frac{W_\bt W^\bt}{32\pi^2}\left(b_0 + \sum_i T(R_i) \gm_i\right)
+ \left(3W-\sum_{i=1}^n
\phi_i\frac{\del W}{\del \phi_i}(1+\hlf\gm_i)\right)\right]
\ee
Assuming the superpotential has the form of a polynomial
\be
W(\phi_i) = \sum_s h_s W^{(s)}(\phi_i)
\ee
where $W^{(s)}$ is a product of $d_s$ fields, we may rewrite the anomaly
as
\be\label{DJ}
\br D^{\dot \al}J_{\al\dot \al}=
-\frac{1}{3}D_\al
\left[ \frac{W_\bt W^\bt}{32\pi^2}\left(b_0 + \sum_i T(R_i) \gm_i\right)
 +\sum_s h_s \left((d_s-3)W^{(s)}
        +\sum_i\hlf\gm_i\phi_i\frac{\del W^{(s)}}{\del \phi_i}\right)\right]
\ee

Thus, to have a theory with no scale dependence, the \batas
\be\label{ganom}
A_g = -\left[3C_2(G) - \sum_i T(R_i) + \sum_i T(R_i) \gm_i \right]
\ee
and, for each $s$,
\be\label{hanom}
A_{h_s}=
(d_s-3) +\hlf\sum_i\gm_i\frac{\del \ln W^{(s)}}{\del \ln \phi_i}
\ee
must vanish.  (The expression $\frac{\del \ln W^{(s)}}{\del \ln \phi_i}$
simply counts the number of times $\phi_i$
appears in $W^{(s)}$.)  Near a fixed point
the $\bt$-functions for the gauge coupling $g$ and the
superpotential couplings $h_s$ must be proportional to these conditions.
In our conventions, as can be read off from the classical and one-loop
behavior of \eqr{DJ}, a coupling
is driven to zero if its \bata\ is positive and away from zero if it
is negative.

The condition for a fixed point is that all \batas\ \eqr{ganom},
\eqr{hanom} vanish.
This puts $n$ constraints on the $n$ couplings $g,\{h_s\}$.
If these constraints are linearly independent, then we expect
their solutions to be isolated points in the space of coupling constants.
But if only $p$ constraints are linearly independent, then
the generic solution to the vanishing of the \batas\ will
be an $n-p$ dimensional manifold of fixed points.  Of course,
it is always possible that the constraints have no solutions, either
because they put contradictory conditions on the anomalous dimensions,
or because there are no values of the coupling constants for which
the anomalous dimensions satisfy them.

\subsection{Discussion}
\label{subsec:discuss}

The exact validity of the formulas \eqr{ganom} and \eqr{hanom},
even non-perturbatively, is of the
utmost importance for us in this paper. Certainly this will be
reliable for those models with lines of fixed points passing through
weak coupling.  Non-perturbative
renormalizations of the superpotential are generally, by holomorphy,
ultraviolet finite, and as long as we are at moderately weak coupling there
should be no strange behavior in the K\"ahler potential which would
invalidate the derivation.  For non-renormalizable effective theories,
however, there are
potential pitfalls.  In Eqs.~\eqr{DJanom} and \eqr{EOM} we implicitly
assumed that the operators we wrote were the most relevant ones.  For this
to be appropriate with a non-renormalizable superpotential, we should
have a substantial gauge coupling at the ultraviolet cutoff (which
of course should not be taken to infinity) so that the superpotential
is nearly marginal there.  This can only happen outside the perturbative
regime.  It is therefore possible for non-renormalizable
operators present in the effective K\"ahler potential to become
marginal or relevant before the
superpotential does.  In this case Eqs.~\eqr{DJanom} and \eqr{EOM}
do not properly characterize the theory.  The appearance
of marginal or relevant non-chiral operators in the action often signals
a breakdown of the description of the theory
in terms of the fields $\phi_i$,
as occurs in \cite{NAD} for gauge group $SU(N_c)$ with $N_f\leq
\frac{3}{2}N_c$ flavors in the fundamental representation.
We will assume that \eqr{ganom} and \eqr{hanom}
are valid in all cases in which the description in terms of the
original fields still makes sense.  The consistency of our results with
other results in the literature \cite{lance}-~\cite{NSKItwo},\cite{NSEW}
suggest that this is so.

At a superconformal fixed point there is an $R$ symmetry which is
part of the superconformal multiplet.  The multiplet contains the
generator of dilations, and as a result the $R$ charge and the dimension
$D_k=d_k+\hlf\gm_k$ of a gauge invariant chiral superfield
field are related\cite{Superconf,Mack}.  In four dimensions
this relation is $D=\frac{3}{2}R$, or $\gm_k=3R_k-2d_k$.  The conditions
\eqr{ganom} and \eqr{hanom} ensure that the $R$ charge is conserved
by the superpotential and has no gauge anomaly.  This is natural,
since the derivative of the supercurrent contains both the scaling
and $R$ anomalies as components. One may rephrase our arguments
for the existence of a marginal operator by searching for a
perturbation of a known fixed point under which the $R$ symmetry is
unambiguously preserved, up to an anomaly-free $U(1)$.
This approach avoids the problems of the derivation presented above, in
that no assumptions are made concerning the K\"ahler potential.
Consequently, we believe that the existence of a stable fixed point which
has a unitary description in terms of the $\phi_i$ probably implies that the
derivation of \eqr{DJ} does not suffer, in the vicinity of that point,
from problems associated with marginal and relevant non-chiral operators.
Furthermore, it is often useful to understand the physics away from but near
the fixed point, and the Shifman-Vainshtein argument gives more
physical insight into the renormalization group flow than does
a discussion limited to the $R$ symmetries at a fixed point.

The operators of the theory form a representation of the
superconformal algebra.  This representation must be unitary.
As shown in \cite{Mack} and employed in \cite{NAD}, this puts
restrictions on the dimensions of operators.  Specifically, the
dimension of any gauge invariant operator must either be zero (in which
case the operator is the identity), one (in which case the operator
is a free field), or greater than one. This will
cause us to discard certain candidate marginal operators because
the theory containing them would have to be non-unitary.
It also ensures that renormalizable models with no gauge invariances
cannot have non-trivial fixed points.  Consider a Wess-Zumino theory
of chiral superfields with a superpotential made from cubic couplings.
By unitarity the superfields must all have dimension one or greater; for the
cubic superpotential to be dimension three, all must have dimension one.
But such fields must be free at a superconformal fixed point, so the
superpotential flows either to zero or to strong coupling where
the description in terms of the original fields breaks down. (In
perturbation theory the anomalous dimensions and the $\bt$-functions
are positive, so the theory is free in the infrared.)
Theories of chiral superfields with non-renormalizable superpotentials
can be rejected as well.  Similarly, in an Abelian gauge theory, the
fact that the gauge field strength superfield
$W_\al=-\frac{1}{4}\br{D}^2 D_\al V$\ is gauge invariant and has
dimension $\frac{3}{2}$ implies that either the theory is free in
the infrared (as one expects perturbatively) or the theory flows
to a region of strong coupling where the description in terms of
the original fields breaks down.  We will therefore consider only
non-Abelian gauge theory for the remainder of the paper.

The equations \eqr{ganom} and \eqr{hanom} are special when
$b_0=0$ and $d_s-3=0$; in this case the \batas\ are homogeneous
in the anomalous dimensions of the fields.  The theory with
zero gauge coupling and no superpotential, for which all anomalous
dimensions vanish, is then
a stable fixed point with a marginal operator.
In some cases the vanishing of the scale anomaly forces the anomalous
dimensions of all fields to vanish, implying the theory has
a manifold of fixed points where its
effective action is finite.  However, in other cases our formalism
does not imply this; the dimensions of some chiral operators, or,
equivalently, the charges of certain fields under the $R$ symmetry in the
supercurrent, are undetermined.  Despite this it
is straightforward to show the anomalous dimensions
always vanish at leading-loop order when Eqs.~\eqr{ganom}-\eqr{hanom}
are satisfied.
At this time we are unable to show that these theories are finite
to all orders, nor are we able to verify the claims of Kazakov
to this effect\cite{finitelate}.  Perhaps finiteness follows
from the fact that these superconformal theories can be continuously deformed
to zero coupling.  The resolution of this issue does not
affect the bulk of our results.

\section{The Fixed Lines of N=4 and N=2 Theories}
\label{sec:SUSYextnd}

The best known models with fixed lines are those of N=4 supersymmetry.
These can be thought of as N=1 theories with a gauge coupling constant
$g$ and three chiral superfields in the adjoint representation
$\Sigma_i$, coupled through the superpotential
$W=h\Sigma_1\Sigma_2\Sigma_3$.  By symmetry, the three fields have
the same anomalous dimension $\gm$.
The N=4 supersymmetry requires $h=g$; but let us relax this condition.
The vanishing of the \batas\
\be
A_g = -3C_2(G)\gm \propto A_h = \frac{3}{2}\gm
\ee
puts only {\it one} condition on {\it two} couplings,
namely $\gm(g,h)=0$.  We know this is true at $g=h=0$, so a
curve of fixed points may pass through the free theory.

  That such a curve exists, at least for weak coupling,
can be easily confirmed from basic physical intuition or by simple
calculation.  For $g\gg h$ the superpotential is negligible
and the theory is an gauge theory which is known to be infrared free;
from \eqr{ganom} this corresponds
to the statement that $\gm(g\gg h)$ is negative.  For $g\ll h$ the
theory is a pure scalar field theory which is infrared trivial and has
a Landau pole, which from \eqr{hanom} tells us that $\gm(g\ll h)$ is
positive.  If $\gm(g,h)$ is continuous, these two regions must be
separated by a curve where $\gm$ vanishes.
The one-loop formula $\gm(g,h)=A(h^2-g^2)$ where $A>0$
confirms this.  A similar argument applies for every weakly
coupled model that we will present.

Furthermore, the behavior of $\gm(g,h)$ on the space
of couplings shows that the fixed line is infrared stable; near but off
the fixed line, the sign of $\gm(g,h)$ is such that the theory is
driven to the fixed line in the infrared.  The fine-tuning
of the couplings which is needed to set $\gm(g,h)=0$ and put the theory
on the fixed line is thus a natural one.  (A similar situation
will be found in all of the models that we study.)
Another way to say this is
that if the N=4 supersymmetry is broken at the ultraviolet cutoff,
it will return as an accidental symmetry in the infrared.
In Fig.~\ref{fig:Nisfour} we illustrate these points.

Finally, we emphasize the simplicity of our arguments.
To show that the curve of fixed points lies on the line
$g=h$ would require the use of the full N=4 supersymmetry.  However,
only N=1 supersymmetry was used in proving the {\it existence} of the
fixed line.  Additionally, the {\it finiteness} of the model, which follows
from $A_g\propto A_h\propto\gm(g,h)=0$, was derived using N=1
supersymmetry alone.

We can see a similar feature in models with N=2 supersymmetry.
Consider a theory with $\nf$ hypermultiplets in some
representation $R$.  Treated as
an N=1 model \cite{}, the matter content is
an adjoint representation $\Sigma$, associated with the gauge fields,
and $\nf$ hypermultiplets consisting of pairs $Q^f,\ot Q_f$
in conjugate representations $R,\br R$.  The superpotential
of the model, $W=h Q^f \Sigma \ot Q_f$, preserves the flavor symmetry,
so we know that all the $Q^f, \ot Q_f$ have the same anomalous
dimension $\gm_Q$.  The \batas\ are
\be\begin{array}{ccc}
A_g=  -(2C_2(G)-\nf T(R))-C_2(G)\gm_\Sigma-\nf T(R)\gm_Q \\
A_h= \hlf( \gm_\Sigma+2\gm_Q)
\end{array}\ee
These are proportional if $b_0=2C_2(G)-\nf T(R)=0$; thus, if the
one-loop gauge $\bt$-function vanishes, we expect a fixed curve
with $g\approx h$.
(A concrete example of the above is the $SU(2)$ model
with $4$ hypermultiplets in the fundamental representation, which was
discussed in Ref.~\cite{NSEW}.)
Again this can be verified perturbatively for small coupling.
Notice that our methods
do not obviously prove finiteness here, since only $\gm_\Sigma+2\gm_Q$
need vanish.  However, by N=2 supersymmetry, the fermion
in the superfield $\Sigma$ must
have the same dimension as the gluino $W_\alpha$, whose $R$ charge is
fixed to be one and whose dimension is therefore its canonical
value of $3/2$.  This ensures that the dimension of $\Sigma$ is its canonical
value, so $\gm_\Sigma$ is zero and thus $\gm_Q$ vanishes as well.

\section{Theories with Weakly Coupled Fixed Lines in N=1 Supersymmetry}
\label{sec:finite}

We now turn to N=1 supersymmetry.  We begin by discussing
models which have fixed lines passing through weak coupling,
many of which have appeared in the literature
\cite{finiteearly}-~\cite{finiteSUfive}.
In \cite{finitelate} and \cite{PandS} these theories are proven to have
marginal operators to all orders in perturbation theory,
though the methods used are quite different from ours.
We believe that our approach is simpler and gives a clearer insight
into the reason for the existence of these models.  Also, to the best
of our knowledge, many of our comments on these examples are original.

As in the finite examples studied above, these theories must have
a gauge coupling with vanishing one-loop $\bt$-function,
so that the origin is a stable fixed point.  The couplings in the
superpotential must therefore be cubic in order that they be marginal
at this point.  Holomorphy and dimensional analysis \cite{Holo} ensure that
these dimensionless couplings are not renormalized by
non-perturbative effects.
Furthermore, in many cases, all anomalous dimensions of
the matter fields necessarily vanish, making the effective
action of these theories finite.  (In cases
where the anomalous dimensions are not constrained to vanish by our
methods, the finiteness of the model is as yet uncertain.)
Specifically, all ultraviolet divergences of the
effective action cancel when we are along the fixed line. This is
not to say however that the theory is divergence-free.  Any field
theory has ultraviolet divergent operator renormalizations; in our
models these appear only for non-chiral operators.
Of course, infrared divergences will be present also.

\subsection{$SU(3)$ with $N_f=9$}
\label{subsec:SUthree}
A simple candidate, first suggested in \cite{finiteearly},
is $SU(3)$ with nine fields $Q^r$ in the fundamental representation
and nine $\aQn{r}$ in the antifundamental. Consider the superpotential
\be\label{suthree}
W=h\left(
\fQn{1}\fQn{2}\fQn{3}+\fQn{4}\fQn{5}\fQn{6}+\fQn{7}\fQn{8}\fQn{9}+
\aQn{1}\aQn{2}\aQn{3}+\aQn{4}\aQn{5}\aQn{6}+\aQn{7}\aQn{8}\aQn{9}\right)
\ .
\ee
Though this superpotential breaks the $[SU(9)]^2\semidirect\ZZ_2$ flavor
symmetry down to $[SU(3)^3\semidirect S_3]^2\semidirect\ZZ_2$,
the $\fQn{r},\aQn{s}$ are still in an irreducible
multiplet of the flavor symmetry and therefore all have the same
anomalous dimension.  The \batas\
$A_g\propto A_h=\frac{3}{2}\gm$
both vanish along the curve $\gm(g,h)=0$, where the theory is
ultraviolet finite.

Strictly speaking, the exactly marginal operator we have found is a linear
combination of \eqr{suthree} and $W^2\equiv{\rm tr} W_\bt W^\bt$,
the square of the gauge chiral superfield, whose
coefficients are determined by the equation
$\gm(g,h)=0$.  This will be true in all the models we present.  Rather
than state this repeatedly, we simply write the matter component
(when it exists) of the marginal operator, leaving its gauge
component implicit.

The importance of maintaining flavor symmetries should not be overlooked.
For example, an operator of the form
$(\fQn{1}\fQn{2}\fQn{3}+\aQn{1}\aQn{2}\aQn{3})$ is not marginal past
leading order.  The matter fields lie in a reducible multiplet
under the $[SU(3)\times SU(6)]^2\semidirect\ZZ_2$ symmetry, so
the anomalous dimensions of the fields
will not all be the same, and the \batas\ will no longer be
proportional.  This can easily be seen  at the one-loop level.

\subsection{$SU(3)\times SU(3)$ with
$3\times[({\bf 3,3})+({\bf\overline3,\overline3})]$}

We now consider a model which has gauge group $SU(3)\times SU(3)$
with three flavors $\fQn{r},\aQn{u}$ in the representations
$({\bf 3,3})$, $({\bf\br3,\br3})$ respectively.
There are two marginal operators
\be\begin{array}{lcr}
h_1 \sum_{r=1}^3\big([\fQn{r}]^3+[\aQn{r}]^3\big) \\
h_2(\fQn{1}\fQn{2}\fQn{3}+\aQn{1}\aQn{2}\aQn{3}) \ .
\end{array}\ee
Since the matter content of the two $SU(3)$ subgroups is the same,
the \batas\ for the gauge couplings are the same; thus a third
marginal operator is a linear combination of the field strength operators
$W_1^2$ and $W_2^2$. The solutions of $\gm(g_1,g_2,h_1,h_2)=0$
form a three-dimensional manifold of fixed points
which passes through zero coupling. This model is similar in appearance
to a two-dimensional Landau-Ginsburg model.  Normally a Landau-Ginsburg
model of chiral superfields in four-dimensions is trivial; it
is a Wess-Zumino type theory of scalars and fermions and the couplings in
its superpotential are known to have positive $\bt$-functions.
However, the addition of gauge interactions to the
model stabilizes it against triviality by reducing the
anomalous dimensions of the matter fields.  Other
models of this type will appear below.

\subsection{$SU(2)\times SU(2)$ with a $({\bf 3,3})$}

This model has an interesting connection with N=4 supersymmetry.
Consider the marginal superpotential $W=hQ^3$.
The one-loop gauge $\bt$-functions are zero; indeed, if we shut
off one of the two
gauge couplings and set $h$ equal to the other, we have an N=4 model.
As in the $SU(3)\times SU(3)$ model above, the vanishing of the three \batas\
occurs along a two-dimensional manifold in $(g_1,g_2,h)$-space; this
manifold passes through
the two N=4 fixed lines at $g_1=h, g_2=0$ and $g_2=h, g_1=0$.
There is a strong-weak coupling duality symmetry along the N=4 lines;
it would be interesting to understand whether
it can be continued in some form (most likely acting
only on the chiral operators of the theory) onto the entire manifold.

\subsection{N=4 with a d-type coupling}
\label{subsec:dtype}

Another finite model \cite{finiteearly,finitelate,listfin} is an $SU(N>2)$
theory with the same matter content
as an N=4 supersymmetric model but with a superpotential that
combines the three adjoint superfields symmetrically rather
than antisymmetrically.  The superpotential
\be\label{fdterms}
W = (h_1\ f^{abc}+h_2\ d^{abc})\Sigma_1^a\Sigma_2^b\Sigma_3^c \ ,
\ee
where $a,b,c$ are adjoint group indices and $f,d$ are the antisymmetric
and symmetric invariants of the group,
has two independent couplings $h_1$ and $h_2$ which have proportional
\batas.   We may also add an independent operator
\be\label{sigcubed}
h_3\ d^{abc}\sum_{i=1}^3  \Sigma_i^a\Sigma_i^b\Sigma_i^c
\ee
Consequently, this N=1 model has a three-dimensional manifold,
of which the N=4 line is a subset, specified by
$\gm(g,h_1,h_2,h_3) = 0$.  Since there is expected to be a duality
relating strong to weak coupling on the N=4 fixed line, it is
tempting to conjecture, as in the previous case,
that this duality extends in some non-trivial way over the entire manifold
of N=1 fixed points.

\subsection{An N=2 model with an additional operator}

The previous example could be considered an N=2 model with two
additional operators.  Certain other finite N=2 models also contain
additional exactly marginal operators.
Consider an $SU(N)$ theory with fields $S^{\{\al\bt\}},\ot S_{\{\al\bt\}}$
in the symmetric tensor representations
and fields  $A^{[\al\bt]},\ot A_{[\al\bt]}$ in the
antisymmetric tensor representations,
along with the adjoint field $\Sigma$ which is part of the N=2
gauge multiplet.  If we consider a superpotential
\be
W = h_A \ot A\Sigma A + h_S \ot S\Sigma S
+ y (\ot S\Sigma A + \ot A\Sigma S)
\ee
it is easy to see that there is a two-dimensional manifold of fixed
points which contains the N=2 fixed line at $y=0,h_A=h_S=g$.
Assuming that this N=2 fixed line has an strong-weak coupling duality
transformation, it would again be interesting  to
understand how it extends to the remainder of this manifold.

\subsection{A chiral example}
\label{subsec:Esix}

A simple chiral model \cite{listfin}
is $E_6$ with twelve fields in the {\bf 27}
representation.  Since three {\bf 27}'s may be combined to form an
invariant, there are marginal operators of the form
\be\begin{array}{l}
h_1{\displaystyle \sum_{r=1}^{12}} (Q^r)^3\\ \\
h_2\left(\fQn{1}\fQn{2}\fQn{3}+\fQn{4}\fQn{5}\fQn{6}
+\fQn{7}\fQn{8}\fQn{9} +\fQn{10}\fQn{11}\fQn{12}\right)
\ .
\end{array}
\ee

A slight modification of this model demonstrates an important point.
If we replace $0<k<12$ of the {\bf 27} fields with ${\bf\br{27}}$ fields,
we no longer have a symmetry (except at $k=6$) which ensures that all
twelve fields have the same anomalous dimensions for arbitary
values of the couplings.  Still, the operator
\be
h\sum_{r=1}^{12-k} (Q^r)^3+h'\sum_{s=1}^k (\ot Q_s)^3
\ee
is marginal.
The \batas\
\be\begin{array}{lcr}
A_g = - (12-k)\gm_Q-k\gm_{\tilde Q}\\
A_h =  \frac{3}{2}\gm_Q\\
A_h' =  \frac{3}{2}\gm_{\tilde Q}
\end{array}\ee
are linearly dependent, so we expect a fixed curve with non-vanishing
and unequal $h$ and $h'$ on which all anomalous dimensions vanish.
(Note that $h$ and $h'$ will be equal at leading orders in perturbation
theory.)  By contrast, the operator with $h\neq 0$, $h'=0$ is not
marginal, since $A_g$ and $A_h$ are linearly independent.

\subsection{An $SU(4)$ model with undetermined $R$ charge}
\label{subsec:sufourQAQ}

A more complicated example is found in an
$SU(4)$ gauge theory with four antisymmetric tensors
$A^{[\al\bt]}_a$, $\ot A_{a[\al\bt]}$ and
eight flavors of fundamentals, $\fQn{r \al}$, $\aQn{r \al}$.
As an illustration of a subtlety, first consider the operator
\be
W=h\sum_{r=1}^{4}\left( \fQn{r} A_1\fQn{r+4} + \fQn{r} A_2\fQn{r+4}
+ \aQn{r}\ot A_1\aQn{r+4} + \aQn{r}\ot A_2\aQn{r+4}\right).
\ee
This is only marginal at leading order.  The fields $A_1$ and $A_2$ will mix
through $Q$ loops, and when the matrix of anomalous dimensions is
diagonalized the combinations $A_1+A_2$ and $A_1-A_2$ will
have different eigenvalues; in fact the latter does not couple
to the $Q$ fields at all.

Instead consider the marginal operator
\be\label{QAQop}
W=h\sum_{i=1}^{2}\left( \fQn{2i-1} A_1\fQn{2i+3} + \fQn{2i} A_2\fQn{2i+4}
+ \aQn{2i-1}\ot A_1\aQn{2i+3} + \aQn{2i}\ot A_2\aQn{2i+4}\right)\ .
\ee
This operator preserves sufficient symmetry to ensure
that the anomalous dimensions of $A$ and $\ot A$ are all equal,
as are those of $Q$ and $\ot Q$. The \batas\ are proportional
\be
\begin{array}{ccc}
A_g= -(4\gm_A+8\gm_Q) \\
A_h= \hlf(\gm_A+2\gm_Q)
\end{array}
\ee
so the theory has a curve of fixed points.  Here we have not shown that
the anomalous dimensions all vanish; only the sum $\gm_A+2\gm_Q$
must be zero on the fixed curve.  It can be shown that all models
of this type have anomalous dimensions which are zero at one-loop,
but finiteness to all orders, in our view, has not been clearly
established\cite{finitelate}.

This situation, where the \batas\ vanish but the anomalous dimensions
are not required to vanish, often arises.  The ambiguity in the
anomalous dimensions is equivalent to the statement that more
than one gauge-anomaly-free $R$ charge is present, and we do not know
which $R$ current appears in the same multiplet as the energy-momentum
tensor.  In this case, in addition to the $R_0$ charge present in the
free theory, there is a charge $X$ with $X(Q)=1$, $X(A)=-2$ which
is anomaly-free, so the dimensions of operators are given by
the charge $R=R_0+\frac{1}{3}X\gm_Q$.  The theory is finite if
and only if $R=R_0$.

\subsection{Other models with undetermined $R$ charge}

There are many other models for which our methods fail to prove
finiteness to all orders, but which do have a fixed curve passing
through zero coupling.   For example, there is a series of models with
$SO(N)$ gauge group ($N=3-10,12,14,18$) with
both spinor and vector representations.  These include a chiral
$SO(10)$ theory \cite{finiteearly} with eight ``generations'' of {\bf 16}'s
and {\bf 10}'s, for which a marginal operator is
 $W=h\sum_{i=1}^8 {\bf 16}_i {\bf 10}_i {\bf 16}_i$.

A more intricate model \cite{finiteSUfive} which was proposed as a GUT
candidate has three generations of matter and Higgs superfields
$\Psi_f,\Lambda_f,H^u_f,H^d_f$ $(f=1,2,3)$ in the ${\bf \br 5,10,5,\br{5}}$
representations, along with a field $\Sigma$ in the
${\bf 24}$ to break $SU(5)$ to the Standard Model gauge group,
and two extra chiral superfields $S,\tilde S$
in the ${\bf 5,\br{5}}$ representations.  Ignoring mass terms,
the couplings in the superpotential are
\be
\sum_{f=1}^3\
\Bigg(
h_1\Big[\Psi_f H^u_f \Lambda_f\Big] +
h_2\Big[\Lambda_f H^d_f \Lambda_f\Big]\Bigg) \ +
h_3 S\tilde S \Sigma + h_4 \Sigma\Sigma\Sigma \ .
\ee
The gauge \bata\
\be
A_g= - 3(\gm_\Psi+3\gm_\Lambda + \gm_{H^u}+ \gm_{H^d})
-\gm_S - \gm_{\tilde S} - 10\gm_\Sigma
\ee
is proportional to $A_1 +A_2+\frac{1}{3}A_3+3A_4$, where
$A_p$ is the \bata\ for $h_p$.

\section{From Weakly to Strongly Coupled Fixed Lines}
\label{sec:weaktostrong}

So far, we have discussed models which were in some sense
generalizations of N=4 and N=2 models in that they contained
trilinear couplings in the superpotential, leading to fixed curves
that passed through zero coupling.  In the remainder of the paper we
will consider models for which, if a manifold of fixed points exists,
it does not pass through the origin.
In this section, we consider a class of models
which interpolates between weakly and strongly coupled manifolds
of fixed points.  The existence of this set of fixed curves suggests
that our approach can be applied to strong coupling fixed points.
We begin with a model which has a weakly coupled large-$N$ limit.

\subsection{$Sp(2N)\times Sp(2N)$ with three  $({\bf 2N},{\bf 2N})$}
\label{subsec:SptimesSp}

Consider a theory with gauge group $Sp(2N)\times Sp(2N)$ with coupling
constants $g_1,g_2$.  The matter fields consist of three multiplets
transforming
in the $({\bf 2N},{\bf 2N})$ representation. We note that if we take
one of the gauge couplings to zero, we obtain an $Sp(2N)$ model
with $6N$ multiplets transforming as ${\bf 2N}$; this model, according to
Ref.~\cite{NAD}, has an interacting fixed point for all $N>1$.  In fact, in
the large $N$ limit, the fixed point is weakly coupled.
Thus for large $N$ we know that there are weakly coupled fixed points
at $g_1=g_*$, $g_2=0$ and at $g_2=g_*$, $g_1=0$, where the matter
fields have anomalous dimension $\gm_*=-N^{-1}$. Furthermore, we expect
there to be a curve of fixed points joining these points, as in
Fig.~\ref{fig:GcrossG},  because, as both $Sp(2N)$ subgroups have
identical matter content, the two \batas\ are equal.
\be\label{btspsp}
A_{g_1} =  A_{g_2} = -3(N+1)+3N -3N\gm(g_1,g_2) .
\ee
Thus, the vanishing of the \batas\ puts only one constraint on the two
coupling constants.  (One can see signs of this fixed curve at one loop;
the anomalous dimension  of the matter fields
is  $\gm\approx-AN(g_1^2+g_2^2)$, $A>0$.)  These results are reliable
for large $N$, but since the proportionality of the \batas\ in \eqr{btspsp} is
true for any $N$, and since it is believed \cite{NAD} that
there are fixed points for $N>1$ where one coupling vanishes, we argue that
fixed curves exist for all $N$.  (It is possible that, for small $N$,
the fixed curves emanating from $g_1=g_*$, $g_2=0$ and $g_2=g_*$, $g_1=0$
are not connected.)  The operator which is exactly
marginal along the fixed curve generates a change in the coupling
constants, so we expect it is a linear combination of $W_{1}^2$
and $W_{2}^2$.  We cannot determine the two coefficients, except at
$g_1=g_2$ where the operator must be $W_1^2-W_2^2$.

\subsection{$G\times G$ with $({\bf R},{\bf R})$ representations}

The existence of the above class of models constitutes evidence that
Eqs.~\eqr{ganom} and \eqr{hanom} can be applied at strong coupling.
We now generalize the previous case to include sets of theories
which do not have a weak coupling extrapolation.
For example, take a theory with any gauge group $G$ and matter
representations ${\bf R_i}$ which
has a non-trivial fixed point.  We conjecture that a
theory with $G\times G$ and matter in
representations $({\bf R_i},{\bf R_i})$ will then have a fixed curve;
the \batas\ are still proportional since
both $G$ subgroups have the same matter content.  Again,
we expect that the marginal operator is something of the form
$W_{1}^2-C(g_1,g_2)W_{2}^2$, with $C(g_1,g_2)$ determined by the dependence
of the anomalous dimensions on the coupling constants.

\subsection{$SU(2)\times SU(2)\approx SO(4)$
with $\nf$ flavors of $({\bf 2},{\bf 2})\approx {\bf 4}$}
\label{subsec:sofour}

The case of $SU(2)\times SU(2)$ with $\nf>3$ copies of
$({\bf 2},{\bf 2})$ is particularly interesting.
According to Ref.~\cite{NAD} this model has interacting fixed points for
$\nf=4,5$.  The arguments of \cite{NSKItwo} and of
this section imply that these have
fixed curves generated by $W_1^2-C(g_1,g_2)W_2^2$.

Furthermore \cite{NSKItwo},
the N=1 duality studied by Seiberg has a number of interesting implications.
N=1 duality \cite{NAD} maps this theory to $SO(\nf)$ with $\nf$
vector multiplets; at $g_1=g_2$ the operator $W_{1}^2-W_{2}^2$
is mapped to the baryon operator of the dual theory $B_D=Q_D^\nf$.
(For $\nf>5$ the dual model is believed not to have a fixed point.)
The fixed curve in the original theory is therefore apparently
mapped to a fixed curve generated by the baryon.  This
suggests that the baryon in the dual theory is exactly marginal.
One may easily confirm that this claim is consistent with the anomaly
coefficients.  On
the other hand, for $\nf=4$, where both the original and dual
theory have gauge group $SO(4)$, the baryon $B=Q^4$ in the
{\it original} theory is mapped to $W_{1D}^2-W_{2D}^2$ in
the {\it dual} theory, which generates
a fixed curve.  In summary \cite{NSKItwo} the case $\nf=4$ has a
two-dimensional manifold of fixed points generated at $g_1=g_2$ by
the exactly marginal operators $W_{1}^2-W_{2}^2$ and $Q^4$.
One may easily check that the \batas\ for the two gauge
couplings and for the baryon operator are proportional.
\be\begin{array}{ccc}
\label{betasoiv}
A_{g_1}= A_{g_2}=  (-12+8)-8\gm_Q\\
A_h=1+2\gm_Q
\end{array}\ee
This puts one condition on the three couplings, showing that
this claim is consistent.

\section{Non-Renormalizable Operators}
\label{sec:nonrenorm}

In the case of $SO(4)$ discussed above an exactly marginal
renormalizable operator was mapped under N=1 duality
to an exactly marginal but perturbatively non-renormalizable operator.
Initially one would have been reluctant to believe
that a formula like \eqr{hanom} would apply to a
non-renormalizable operator.
However, the arguments of Sec.~\ref{subsec:derive} do not require that
the theory have an ultraviolet fixed point, and apply to effective
field theories.  Still, they
can break down as discussed in Sec.~\ref{subsec:discuss}, and one might
have worried that this would always happen.
The existence of  the N=1 duality map and the examples in
Sec.~\ref{subsec:sofour} suggests that, at least in some cases,
the formulas \eqr{ganom} and \eqr{hanom} are appropriate to describe the
physics.

Drawing confidence from the known examples, we now present
other theories which have
candidate marginal operators that are perturbatively non-renormalizable.
In many models, more than one such operator can be found;
we content ourselves with displaying a single example.

\subsection{Some important issues}
\label{subsec:NRissues}

First, we must stress that in all of the cases presented below,
the existence of strongly coupled manifolds of fixed points is merely
conjecture, both because the dynamics of specific non-renormalizable
theories might be more complex than or quite different from what we suppose,
invalidating the arguments of Sec.~\ref{subsec:derive}, and because
some of the models we present may not have any fixed points at all.
Some of these theories were studied in \cite{NAD} and
arguments for fixed points were given; in these
cases we have more confidence that the fixed curves that we conjecture
are really present. Additional support for our approach stems
from certain two-dimensional models where similar phenomena are
known to occur\cite{MikeD}.

The fact that these operators are not renormalizable should not
by itself be cause for discarding our approach, as such theories
do make sense as long as they are considered effective theories
valid below some cutoff $M_0$.   As we will
see later, many of these non-renormalizable field theories
are the low-energy effective expression of a
renormalizable field theory, which may serve as a supersymmetric
ultraviolet cutoff.  This cutoff should not be taken to infinity.
The non-renormalizable terms in the superpotential are
irrelevant when the gauge coupling is weak, so any fixed point
at which they are marginal must have a substantial gauge coupling.
Since the renormalization group flow increases the
gauge coupling logarithmically while suppressing the non-renormalizable
couplings by powers, the scale  at which the theory reaches its
infrared fixed point should not lie too far below $M_0$ and the gauge
coupling at $M_0$ should be finite.

In contrast to the renormalizable theories considered earlier, these
models are subject to non-perturbative renormalizations.  Suppose
the superpotential has coupling constants $h_s$ of negative
dimension.  In general
there is a combination $H$ of the $h_s$ which, when
multiplied by the dynamical scale of the theory $\Lam^{b_0}$,
is dimensionless  and invariant under all global symmetries.
The couplings $h_s$ may then be multiplicatively renormalized
by a holomorphic function $f(H\Lam^{b_0})$.
The function $f$ may have singularities of various types, so
the renormalized couplings $h^R_s = h_s f(H\Lam^{b_0})$ may not be
finite for all finite values of the bare couplings $h_s$.
However, if any symmetries
are restored at $h_s=0$ (as is always the case in our examples)
then $h_s^R$ must be zero there also, from which it follows that
the function $h_s^R$ must be finite as $h_s=0$ is approached along any
direction.  We therefore conclude that although $h_s^R$ may be
infinite for particular values of $h_s$, there is a neighborhood
of the point $h_s=0$ where the renormalized couplings are
finite and the existence of a marginal operator with marginal coupling
$h_s^R$ may be established using our methods.

Almost all of the non-renormalizable superpotentials we study involve
operators of dimension four or five.  An occasional dimension six
operator may arise.  Beyond dimension six it is difficult to find a
marginal operator.  The reason has to do with unitarity
and the dimensions of gauge invariant operators \cite{Mack};  at a
superconformal fixed point
the dimensions of all non-trivial gauge invariant operators must be greater
than or equal to one.  If an operator of canonical dimension $d$ is to be
marginal, then, at the fixed point, the dimension
of at least one of the fundamental fields
which it contains must be $d_\phi\leq 3/d$, which is less than $\hlf$
if $d>6$.  It is difficult (though probably not impossible)
to construct a situation where $d_\phi<\hlf$ for some field $\phi$,
yet there are no gauge invariant bilinears of dimension less than
one.  In particular this can only happen in a chiral theory, since in
a vector-like theory there will be a field $\tilde\phi$ (which may be
$\phi$ itself) with $d_{\tilde\phi}=d_\phi$, giving a meson $\phi\tilde\phi$
of dimension less than one.

Even in the case of a candidate
marginal operator of dimension six, there may be difficulties in a
vector-like theory, since the
meson $\phi\tilde\phi$ has dimension one and must therefore be free.
Our methods often cannot determine whether or not a given operator
is marginal in this case.  An example which has a
dimension six marginal operator will be given in Sec.~\ref{subsec:SUsix}.
As an example where a dimension six operator satisfies our conditions
but fails to create a marginal deformation, consider a
theory with gauge group $SO(6)$ and six vector representations. This theory
has a candidate marginal baryon operator, which is mapped under N=1
duality \cite{NAD,NSKItwo} to the operator $W_1^2-W_2^2$ in an
$SO(4)\approx SU(2)\times SU(2)$ theory with six vector representations.
When this operator is turned off, the $SO(6)$ theory flows to strong
coupling, while the $SO(4)$ model is free in the infrared \cite{NAD}.
The operator $W_1^2-W_2^2$ thus has no physical
consequences in the $SO(4)$ theory, leading us to suspect that the marginal
baryon operator in $SO(6)$ is an irrelevant perturbation of the
strongly coupled theory.

\subsection{Some $SU(4)$ examples}
\label{subsec:sufours}

Amongst the $SU(\nc)$ interacting fixed points studied in \cite{NAD}, we find
one example $(\nc=4,\nf=8)$ with a marginal baryon operator.
As in the $SU(3)$ example considered earlier, we need an operator
which is marginal at the known infrared fixed point and
which preserves enough flavor symmetry that all fields have
the same anomalous dimension $\gm_Q$.   A suitable choice is
\be\label{sufourB}
 h(\fQn{1}\fQn{2}\fQn{3}\fQn{4}+\fQn{5}\fQn{6}\fQn{7}\fQn{8}+
      \aQn{1}\aQn{2}\aQn{3}\aQn{4}+\aQn{5}\aQn{6}\aQn{7}\aQn{8}).
\ee
Both the gauge and Yukawa \batas\ are proportional to $1+2\gm_Q$.
The qualitative features of the renormalization group flow are
illustrated in Fig.~\ref{fig:SUiv}.

Up to now we have ignored the differences
between the original and dual models under N=1 duality \cite{NAD}.  In
particular we have disregarded the fact that the dual model
contains singlets and a superpotential coupling them to the other
fields.  It is useful to focus our attention briefly
on this point, though it will be seen that the effect of the
singlets is minimal.
In the case at hand the dual theory \cite{NAD} has the same color
and flavor content as the original,
though the flavor representation of the fields is conjugate to that
of the original theory.  In addition to the colored fields $\fq,\aq$
there are color-singlet fields \Miti\ which are coupled
through the superpotential $W=\lam\Mitim\fq\aq$.
 From the operator mapping described
in \cite{NAD} we expect the marginal operator
to be the same as the above with $Q$ replaced by $q$.
The \batas\ for the three coupling constants $g,h,\lam$ are
\be\begin{array}{ccc}
A_g= - 4 - 8\gm_q(g,h,\lam) \\
A_h = 1 +2 \gm_q(g,h,\lam) \\
A_\lam=\hlf\gm_M(g,h,\lam)+\gm_q(g,h,\lam) \ .
\end{array}\ee
Since the first two are proportional (in fact they are the same
as in the original model)  there are two constraints on
three couplings, and we expect a fixed curve emanating from the
conjectured fixed point at $(g,h,\lam)=(g_*,0,\lam_*)$.  The
coefficient $A_\lam$ simply sets $\gm_M=1$ on the fixed curve and
is otherwise inert.  We also learn that $\lam_*=0$ would not be a
stable fixed point. At $\lam=0$ the singlets
decouple from the theory and, were this a fixed point, their anomalous
dimensions would have to vanish by unitarity.  As a result
$A_\lam$ would be negative
(since $\gm_q\approx-\hlf$ there) so the theory would be driven
away from $\lam=0$.  We conclude that $\lam_*\neq 0$.

Another very similar model has
eight antisymmetric tensors; an exactly marginal operator is
\be
 h \eps_{\al\gm\theta\kappa}\eps_{\bt\delta\eta\lam}
(A_1^{\al\bt} A_2^{\gm\delta} A_3^{\theta\eta} A_4^{\kappa\lam}
+ A_5^{\al\bt} A_6^{\gm\delta} A_7^{\theta\eta} A_8^{\kappa\lam})\ .
\ee
Since this model is equivalent to $SO(6)$ with eight vector
multiplets, which has a fixed point at $h=0$ \cite{NAD,NSKItwo},
we are confident that it has a fixed curve.

To illustrate the issues associated with models which lack a unique
$R$ charge, we present two more $SU(4)$ examples.  The first
has four $({\bf 4}+ {\bf\br 4})$ representations
$\fQn{r},\aQn{}^r$ and four ${\bf 6}$
representations $A^r_{[\al\bt]}$; a suitable operator is
\be\label{QAAQop}
 h\sum_{r=1}^4 \aQn{}^r A^r A^r \fQn{r} \ .
\ee
In this case the dimensions of operators are not determined.
However, by adding another marginal operator
\be
y_Q(\fQn{1}\fQn{2}\fQn{3}\fQn{4}+   \aQn{1}\aQn{2}\aQn{3}\aQn{4})
+ y_A {\rm tr} (A^1\ot A_1 A^2\ot A_2 + A^3\ot A_3 A^4\ot A_4)
\ee
we can fix the dimensions of the fields.  This is reminiscent
of the situation in two-dimensional Landau-Ginsburg models.

Finally, consider a theory
with fields $\fQn{},\aQn{},A,\tilde A,\Sigma$ in
the ${\bf 4},{\bf\br 4},\bf 6,\bf 6,15$ representations.
A candidate operator is
\be\label{sufourSAQ}
h \aQn{}\Sigma\Sigma Q + h' \tilde A\Sigma\Sigma\Sigma A
\ee
The \batas\
\be\begin{array}{lcr}
A_g = -[ 5+4\gm_\Sigma +\gm_A+\gm_{\tilde A}+\hlf\gm_Q+\hlf\gm_{\aQn{}}]\\
A_h =  1 +\gm_\Sigma + \hlf\gm_Q+\hlf\gm_{\aQn{}} \\
A_{h'} =  2 +\frac{3}{2}\gm_\Sigma + \hlf\gm_A+\hlf\gm_{\tilde A}
\end{array}\ee
are linearly dependent.  Unlike the previous case there is no marginal
superpotential which can fix the dimensions of the fields.

\subsection{$SU(6)$ with nine $({\bf 6}+ {\bf\br 6})$ and singlets}
\label{subsec:SUsix}

The N=1 duality of Ref.~\cite{NAD} has interesting implications for
this model, which is dual to that of Sec.~\ref{subsec:SUthree}.
The superpotential
\be\label{susix}
\lam M^r_s  q_r\ot q^s +
h( B^{123}+B^{456}+B^{789}+\ot B_{123}+\ot B_{456}+\ot B_{789}) \ ,
\ee
where $B^{123}\propto q_4q_5q_6q_7q_8q_9$, {\it etc}., is
a candidate marginal operator.
According to Ref.~\cite{NAD}, when $h=0$ this model runs to infinite coupling
and does not reach a fixed point.  This corresponds
to the fact that its dual, $SU(3)$ with nine flavors, runs
to zero coupling.  However, we showed earlier that in fact
its dual has a fixed curve
generated by the operator \eqr{suthree}, which is dual \cite{NAD} to
the operator \eqr{susix}.  It is therefore tempting to suggest
that when $h\neq 0$ the $SU(6)$ model does not run all the way to
infinite coupling, stopping instead on a fixed curve which
runs off to infinite $g$ as $h\rarr 0$, as in Fig.~\ref{fig:SUvi}.
The fact that the meson operator
$q^r\ot q_s$ has dimension one does not rule out this possibility,
as it is redundant. If \eqr{susix} is indeed a marginal operator, it
is possible that semi-classical
methods applied on the fixed curve of its weakly coupled dual can be
used to study the N=1 duality transformation.

\subsection{A Chiral Model}

The strongly interacting models we have considered up to now
are vector-like.  Here we present a candidate chiral model:
$SU(5)$ with five generations $A_i,\ot Q_i$
of ${\bf 10}+{\bf\br 5}$. Using ${\bf\br 5}\in{\bf 10}\times{\bf 10}$
(symmetric combination) and ${\bf 5}\in{\bf 10}\times{\bf \br 5}$,
one can see there is an operator
\be
h\sum_{i=1}^5 A_i A_i A_i\ot Q_i
\ee
which maintains the five-fold flavor symmetry.
The index of the $Q$ field is one while that of $A$ is three; thus the
\batas\ are proportional
\be\begin{array}{ccc}
A_g=  -5 - \frac{15}{2}\gm_A-\frac{5}{2}\gm_Q\\
A_h= 1+\frac{3}{2}\gm_A+\frac{1}{2}\gm_Q
\end{array}\ee
and this model can have a fixed curve.

\section{Renormalization group flow between fixed manifolds}
\label{sec:RGflow}

In this section we study flow between fixed manifolds.  In
particular we would like to understand what happens when
a field is given a mass and integrated out, or when the Higgs
mechanism breaks part of the gauge symmetry.  We will focus
largely on the first case, showing first that manifolds
of fixed points often flow to new ones when a theory is perturbed
by a mass term, and then giving a number of examples.  Finite
theories with N=2 supersymmetry flow to a wide variety of
interesting models.  These include certain special cases studied
in \cite{NAD}; we identify certain suggestive properties
of these models that seem to relate N=2 duality to N=1 duality.
(We will return to this issue in Sec.~\ref{sec:duality}.)
Finally we mention a couple of examples in which symmetry breaking
causes flow from one fixed manifold to another.

\subsection{Effect of integrating out a field}
\label{subsec:IOtheorem}

We will now prove the following lemma.
Consider a theory at a fixed point with a  marginal operator
given by a polynomial superpotential
\be
W=\sum_s h_s W^{(s)}
\ee
where each term $W^{(s)}$ is a gauge-invariant product of at
least three fields.
Suppose a mass term may be written for two of the fields $\phi$, $\phi'$
(not necessarily different.)  If each term $W^{(s)}$ contains either one factor
of $\phi$ or one factor of $\phi'$, then,
when the mass term $m\phi\phi'$ is added as
a perturbation on the fixed point theory, either the theory does not
find a new fixed point, or, if it does, that fixed point has a candidate
marginal operator.

To prove this is straightforward.  For simplicity we consider a
superpotential with only two terms,
of the form $W =  h\phi X + h'\phi' X'$,
where  $X$ and $X'$ are multi-linear in superfields but contain
neither $\phi$ nor $\phi'$.  (From this specific example, the proof
is easily extended to the general case.
We omit the details as they would generate more notation than insight.)

Consider the addition of the mass term $m\phi\phi'$
to the superpotential.
When we integrate out these fields we should implement the equations
\be\label{IOphis}\begin{array}{lcr}
\frac{\del W}{\del \phi} =hX= - m\phi'\\
 \frac{\del W}{\del \phi'}= h'X'= - m\phi \
\end{array}\ee
as operator statements.
The left-hand sides of these equations are by assumption independent
of both $\phi$ and $\phi'$.  This leads to a new superpotential
\be
W^{{\rm new}} = - \frac{hh'}{m} XX'
\ee
We will now show
that this theory has at least one candidate marginal operator.

Since the original theory had a marginal operator, we know that there
is some linear combination of the \batas\ which is zero, which we
may write as
\be\label{oldscs}
c A_{h} +c' A_{h'}
= A_g = -[b_0 + \sum_i T(R_i)\gm_i]
\ee
The form of the superpotential and Eq.~\eqr{hanom} imply that
$A_h = \hlf\gm_\phi+\cdots$, $A_h' = \hlf\gm_{\phi'}+\cdots$.
 From the previous equation and the existence of the mass term $m\phi\phi'$,
which implies $T(R_\phi)=T(R_{\phi'})$, it follows that $c=c'=-2T(R_\phi)$.
The new interaction is of the form $HXX'$
where $H=-h h' /m$. Its \bata\ is
\be\label{sumC}
A_{H} = 1+ (A_{h}-\hlf \gm_\phi) + (A_{h'}-\hlf \gm_{\phi'})
\ee
Using Eqs.~\eqr{oldscs}--\eqr{sumC},
\be
c A_{H} = A_g - T(R_\phi)
(1-\gm_\phi) - T(R_{\phi'})(1-\gm_{\phi'}) \ .
\ee
Following the integrating out of $\phi$ and $\phi'$, the new gauge \bata\
is
\be
A_g^{new} = A_g -  T(R_\phi)(1-\gm_\phi)- T(R_{\phi'})(1-\gm_{\phi'}) =
 c A_{H}
\ee
which indicates that the new superpotential contains one or more candidate
marginal operators.

The new \batas\ put some constraints on the anomalous dimensions
which must be satisfied at a new fixed point.  We could have gotten
the same constraints in another way.  The mass term $m\phi\phi'$
broke the $R$ symmetry which was satisfied at the old fixed point,
but it conserves a new $R$ symmetry.  The linearity of the
superpotential in $\phi$ and $\phi'$ ensures both that a new $R$
exists and that unitarity is not violated by the presence of
zero or negative dimension fields.  Requiring the new $R$ charge to
be conserved classically and quantum mechanically puts conditions
on the anomalous dimensions which are the same as those from
the anomaly coefficients.  One may see this
more clearly by keeping the fields $\phi$ and $\phi'$ in the theory,
treating the mass term as a new interaction,
and rederiving the lemma.  Once at the new fixed point,
integrating out the fields $\phi$ and $\phi'$
does not change the conserved $R$ charge or the dimensions of
chiral operators.  In particular, an operator is marginal whether expressed
in terms of the fields $\phi$ and $\phi'$ or in terms of the
other fields using \eqr{IOphis}.  This will be important later.

We note also that the lemma applies to a wider class of superpotentials.
If, to the superpotential $W$ above, we add terms $W'$ that are independent
of $\phi$ and $\phi'$, then the low energy superpotential will
be $W^{{\rm new}}+W'$.  If $W+W'$ has a marginal operator, and
the fields in $W'$ can be assigned new $R$ charges such that
$W^{{\rm new}}+W'$ has a conserved $R$ symmetry, then the
low-energy theory will have a candidate marginal operator.

Before applying these ideas we make some general observations.
A candidate operator may fail to be marginal under various
circumstances---if it is a composite of redundant operators,
if it preserves an $R$
symmetry which does not characterize any fixed point of the theory,
or if its canonical dimension is too high, preventing the putative
fixed point from being unitary.
The last of these reasons implies that the process of integrating
out fields will terminate fairly quickly, as unitarity comes
into question when dimension six operators begin to appear.

When we derive a non-renormalizable superpotential from a
finite theory, the latter serves as an ultraviolet regulator for
the former, ensuring that non-renormalizability of the effective
theory does not generate
uncontrollable infinities while leaving intact the infrared properties
which we are studying.   The existence of a sensible regulator should
in our view allow the arguments of Sec.~\ref{subsec:derive} to be applied
near a previously established fixed point, subject to the limitations
described in Sec.~\ref{subsec:discuss}.

\subsection{Application to $SU(3)$ with $N_f=9$}

We may immediately find some new non-renormalizable candidate
operators by turning to some finite models and integrating out
selected fields.  For example, if we take the model of
Sec.~\ref{subsec:SUthree} and add a mass term $m\fQn{9}\aQn{9}$,
we generate a theory with eight flavors and superpotential
\be
W=h\left(
\fQn{1}\fQn{2}\fQn{3}+\fQn{4}\fQn{5}\fQn{6}+
\aQn{1}\aQn{2}\aQn{3}+\aQn{4}\aQn{5}\aQn{6}\right)
+H(\fQn{7}\fQn{8}\aQn{7}\aQn{8})
\ .
\ee
Writing $\gm_Q$ for the first six flavors and $\gm_{\hat Q}$ for the
last two, the \batas\ are
\be\begin{array}{lcr}
A_g= -[ 1 + 6\gm_Q + 2\gm_{\hat Q}] \\
A_h=   \frac{3}{2}\gm_Q \\
A_H=  1 + 2\gm_{\hat Q}
\end{array}\ee
which shows that this superpotential is a candidate marginal operator.
However, the flavor symmetry of this model does not treat the eight
quarks symmetrically --- the anomalous dimensions $\gm_Q=0$ and
$\gm_{\hat Q}=-\hlf$ are not equal ---
and thus this fixed curve cannot be continued
to $h=H=0$, where the full $SU(8)\times SU(8)$ flavor symmetry would be
discontinuously restored.  Therefore, if a fixed curve is generated
by this operator, it does {\it not} pass through the fixed point
found in \cite{NAD} in the absence of a superpotential.  Of course
there simply may be no fixed curve of this type.

Now let us add $m(\fQn{3}\aQn{3}+\fQn{6}\aQn{6}+\fQn{9}\aQn{9})$.
The resulting superpotential
\be\label{ffaaop}
W=H(\fQn{1}\fQn{2}\aQn{1}\aQn{2}+ \fQn{4}\fQn{5}\aQn{4}\aQn{5}+
\fQn{7}\fQn{8}\aQn{7}\aQn{8})
\ee
is obviously a candidate marginal operator.  Furthermore, this operator
does preserve the symmetry among all six flavors and can
generate a marginal
deformation of the fixed point with $N_c=3, N_f=6$
studied in \cite{NAD}. Thus, unlike the previous case, there is some
evidence for the existence of a fixed curve in this model.

\subsection{Application to a chiral model}

We may find similar operators in some chiral theories.  Taking
as a starting point the finite $SO(10)$ model with eight ``generations''
of ${\bf 16}+ {\bf 10}$ and marginal operator
$h\sum_{i=1}^8 {\bf 16}_i {\bf 10}_i {\bf 16}_i$, we may integrate
out all the {\bf 10} fields.  Different operators emerge
depending on how this is done.  A diagonal mass term
$m\sum_{i=1}^8 {\bf 10}_i {\bf 10}_i$ generates
a candidate marginal operator
\be
H\sum_{i=1}^8 {\bf 16}_i  {\bf 16}_i {\bf 16}_i  {\bf 16}_i
\ee

\subsection{Application to an $SU(4)$ model}

In Sec.~\ref{subsec:sufourQAQ} we studied a model of four antisymmetric
tensors and eight flavors in the fundamental representation;
this was a finite model whose marginal operator was given in \eqr{QAQop}.
Mass terms for the antisymmetric tensor fields lead to a theory
with eight flavors, studied in \cite{NAD} and in
Sec.~\ref{subsec:sufours}, that has various
operators, of the form $Q^4,(Q\ot Q)^2,\ot Q^4$, which can be assembled
into exactly marginal combinations such as Eq.~\eqr{sufourB}.
Mass terms $m\fQn{r}\aQn{r}$ for $r=5,6,7,8$ lead to another theory
studied in Sec.~\ref{subsec:sufours}; the resulting superpotential
\be
H\sum_{i=1}^{2}\left( \fQn{2i-1} A_1\ot A_1\aQn{2i-1}
                      + \fQn{2i} A_2\ot A_2\aQn{2i}\right)\
\ee
differs only slightly from the operator presented in Eq.~\eqr{QAAQop}.

\subsection{Application to finite N=2 models}
\label{subsec:IONN}

The applications to N=2 models are wide-ranging and interesting.
Finite N=2 models with a single type of group representation
were discussed in Sec.~\ref{sec:SUSYextnd}.  Here we reconsider these along
with theories involving more complicated matter representations.
Since many or all of these N=2 models may have some form of duality,
the N=1 models they flow to may also display duality of some type.
In particular we will see evidence below that the N=1 duality of \cite{NAD}
is related to N=2 duality.  This will be further explored in
Sec.~\ref{sec:duality}.

First consider the model of Sec.~\ref{subsec:dtype}, which has the
matter content of an N=4 model and three marginal operators.
When we add a mass term $m(\Sigma_3)^2$ and
integrate out one of the superfields in the presence
of the superpotential \eqr{fdterms} we find a low-energy theory with
two adjoint superfields $\Sigma_1$ and $\Sigma_2$ and a superpotential
made up of the three candidate marginal operators
\be
(H_1 f^{abc}f^{ade}+H_2 f^{abc}d^{ade}+ H_3 d^{abc}d^{ade})
\Sigma_1^b\Sigma_2^c\Sigma_1^d\Sigma_2^e
\ee
If we add the operator \eqr{sigcubed} to the high-energy superpotential,
then the lemma of Sec.~\ref{subsec:IOtheorem} does not apply.  In
particular there is no conserved $R$ charge when the mass term is added,
so no fixed point can be reached until certain
couplings have flowed to zero.

Next consider a finite N=2 model of $SU(4)$ with three anti-symmetric tensor
representations $(A^i,\ot A_i)$ and two fundamental representations
$(\fQn{r},\aQn{r})$.  To find a candidate marginal operator we may either add
a mass term $m\ {\rm tr} \Sigma^2$ or some combination of masses for
the matter fields.  A variety of different theories result.
For example, if we add
\be
m\left(\ot A_1 A_2+ \ot A_2 A_3+ \aQn{1}\fQn{2}\right)
\ee
we arrive at the last model in Sec.~\ref{subsec:sufours}
and its candidate marginal operator \eqr{sufourSAQ}.

An interesting set of theories are the finite N=2 models
involving only hypermultiplets in the defining representation.
For example, consider $SU(N)$ with $2N$ hypermultiplets $(\fQn{r},\aQn{r})$.
We may add mass terms of a number of types.   By giving
flavor off-diagonal masses to some of the hypermultiplets,
one arrives at superpotentials which are a sum of operators
such as $\ot Q (\Sigma)^k Q$; special cases are those
where all but $N/p$ of the hypermultiplets are integrated out,
for which the superpotential consists only of terms of the form
$\ot Q (\Sigma)^p Q$.  Of course unitarity is lost for $p>4$.

The most intriguing option, however, is to give a mass $m_\Sigma$
to the adjoint field $\Sigma$ and integrate it out.
We then arrive at a model which was studied
in Ref.~\cite{NAD}, $SU(\nc)$ with $2\nc$ flavors, along with a
superpotential
\be\label{QQQQop}
W=
-h(Q^r_\al \ot Q_r^\bt)\  T^{a\al}_\bt T^{a\gm}_\delta \
(Q^s_\gm \ot Q_s^\delta) =
 -\frac{h}{2}\left[(Q^r_\al \ot Q_s^\al)(Q^s_\bt \ot Q_r^\bt)
-\frac{1}{\nc} (Q^r_\al \ot Q_r^\al)(Q^s_\bt \ot Q_s^\bt)\right]
\ee
where $T^a$ are color group matrices. We have used an $SU(\nc)$
group identity to rewrite the superpotential in terms of gauge singlets.
This operator is exactly marginal.  Having written
it in this form it is natural to rewrite it using an auxiliary
meson field $N_r^s$ of canonical dimension two.\footnote{We thank
K.~Intriligator for a discussion which led to this line of thought.}
\be\label{MMop}
W= \lam N_r^s(Q^r_\al \ot Q_s^\al) +
        \frac{\lam^2}{2h} \left[
N_r^s N_s^r-\frac{1}{\nc}N_r^r N_s^s\right]
\ee
Again, all we have done here is rewrite the operator \eqr{QQQQop}
in a new way.  (A similar mechanism has been used in \cite{NSKItwo}.)
As $h\rightarrow 0$, the coefficient of the
marginal operator \eqr{QQQQop} vanishes and the theory reduces
to the original $\nf=2\nc$ model
studied in \cite{NAD}.  As $h\rightarrow \infty$, the meson
mass term in \eqr{MMop} vanishes and the theory becomes remarkably similar
to the N=1 dual \cite{NAD} of the $\nf=2\nc$ model.  It seems that
the fixed curve generated by the operator
\eqr{QQQQop}, \eqr{MMop} connects one theory to the other.

Since the small $h$ theory is found by taking the gauge coupling
in the N=2 theory to be small, we may expect that the when the N=2
gauge coupling is large the low energy theory is in the large $h$
region.  It is natural
to suggest that the weak-strong coupling duality (S-duality)
of the N=2 theory translates into some form of duality
in the N=1 theory.

This naive picture is far too simplistic, and we provide a
more careful though still incomplete analysis of the situation
in Sec.~\ref{sec:duality}.  At this stage we simply note that
the same approach can be used for $SO(N)$ and $Sp(2N)$
gauge groups; the insertion of the auxiliary meson field allows one
to rewrite the marginal operator in a form which suggests that
the curve of fixed points that it generates connects
the low-energy theory to its N=1 dual, or at least to a theory
very similar to its dual.

\subsection{Flow under symmetry breaking}
\label{subsec:symbreak}

Because of the changes in the
representation content of a theory when gauge symmetries are broken,
including the appearance of gauge singlet fields which we have
largely avoided in this paper, the renormalization group flow associated
with symmetry breaking is a rather complicated subject.
It deserves a more systematic study than we have given it
here.   Under many circumstances, symmetry
breaking leads a theory with a marginal operator to flow to a
theory with a marginal operator of lower dimension.  In general this
occurs, as in the case of integrating out a massive field, when
the symmetry breaking preserves a unitary and anomaly-free
$R$ charge consistent with the new flavor symmetries.

A first example involves the $SO(N)$ models with $N$ vector
representations, studied in Sec.~\ref{subsec:sofour}.  The baryon in $SO(5)$
with five flavors generates a curve of fixed points.
As noted in \cite{NSKItwo},
if we break the gauge symmetry to $SO(4)$ and then to $SO(3)$
this fixed curve flows to the fixed curves
generated by the baryons of $SO(4)$ and $SO(3)$ respectively. The
latter theory is N=4 supersymmetric and its fixed line passes
through zero coupling.

Another more complicated example involves $SU(3N)$ with $2N$ flavors
and an adjoint field.  The superpotential
\be\label{QQSQQ}
h\sum_{r=1}^{2N} \fQn{r} \Sigma\Sigma\Sigma\ \aQn{r}
\ee
is a candidate marginal operator.  As discussed in the
previous section, the finite N=2 model with
gauge group $SU(3N)$ and $6N$ flavors flows to this model
when $4N$ of the flavors are integrated out.  Now give vacuum expectation
values to $\fQn{r}\aQn{r}$ for $r=1,\dots,2N$.  This breaks the
gauge symmetry to $SU(N)$.  The fields $\fQn{r},\aQn{r}$ are
eaten by the broken gauge fields, except for some neutral Higgs bosons.
The field $\Sigma$ transforms under the unbroken $SU(N)$
symmetry as
\be
{\bf\rm adjoint}\rightarrow {\bf\rm adjoint}
+ \ {\bf N}\times 2N \ + \ {\bf\br N}\times 2N \ + \ {\bf 1}\times 4N^2 \
\ee
Thus there are a number of singlet fields, along with the matter
content of a finite N=2 $SU(N)$ model.  A number of couplings appear in
the superpotential.  The conditions for a marginal operator have a unique
solution, where all anomalous dimensions vanish.  From unitarity
considerations, a gauge
singlet with vanishing anomalous dimension must decouple, so
all couplings involving singlets flow to zero.  The remaining
superpotential is that of an N=2 theory along with a $y\Sigma^3$
interaction.  From the \batas\ there can be only a one-dimensional
manifold of fixed points; we therefore expect $y$ to flow
to zero and the theory to be driven onto the N=2 fixed line. From a
one-loop computation one may confirm that this is the
only possibility at weak coupling.  Thus, one can flow
from an N=2 fixed line to the fixed curve of this model by integrating
out fields, and from this model down to another N=2 fixed line by
breaking gauge symmetries.

Finally, we note that the example of Sec.~\ref{subsec:SUsix} has
the property that when the gauge symmetry is broken to $SU(3)$,
leaving six flavors, the theory is related under N=1 duality
to a model with the same color and flavor groups but with no singlet
fields.  We have seen
that the theory without singlets has a marginal operator \eqr{ffaaop}.
In the broken $SU(6)$ model, the dual of the operator \eqr{ffaaop}
\be\label{ffaaopD}
(M^1_1 M^2_2-M^1_2 M^2_1) + (M^4_4 M^5_5-M^4_5 M^5_4)
+ (M^7_7 M^8_8-M^7_8 M^8_7)
\ee
is generated by instanton effects \cite{RLMS}.  One may easily check
that when combined with the classical superpotential
$W=M^r_s q_r\ot q^s$ it represents a marginal operator in this
description.

\section{Meson mass operators and N=1 duality}
\label{sec:duality}

In Sec.~\ref{subsec:IONN} we showed that there was a tantalizing
connection between certain finite N=2 models
and those N=1 theories studied by Seiberg \cite{NAD}
which have the same gauge group in both the original and dual descriptions.
(A suggestion that such a connection might exist was made in \cite{NAD}
and related connections are present in \cite{NSKItwo}.)  We considered
a finite N=2 model with gauge group $SU(\nc)$ and
$2\nc$ hypermultiplets in the fundamental representation.  We added
a mass $m_\Sigma$ for the adjoint chiral field, leading to a low-energy
theory that we will call the \djn model.  We observed that the fixed line
of the N=2 model flows to the infrared fixed curve of the \djn
model.  Examining the superpotential \eqr{QQQQop} and its alternate
form \eqr{MMop}, we noted the similarity of certain limits of the \djn
model to the original and dual
$SU(\nc)$ theories with $\nf=2\nc$ studied in \cite{NAD}.

In this section
we present speculative but, we hope, plausible arguments that the N=1
duality of \cite{NAD} is closely associated with S-duality of finite
N=2 models.  We will find a theory, which we will call supersymmetric
quantum chromesodynamics (SQCMD), that has
the same infrared fixed curve as the \djn model.  We will learn more about
the fixed curve of \djn by studying SQCMD in detail.  Eventually
we will use conjectures about N=2 duality to guess its relationship
to N=1 duality.  We cannot verify our guess because we do not
at this time have enough information about duality in the relevant
N=2 theories.

\subsection{Supersymmetric quantum chromesodynamics (SQCMD)}

Consider a theory of $SU(\nc)$ with $2\nc$ flavors $\fQn{r},\aQn{s}$
in the $\bf N_c$ and $\bf\br N_c$ of color, along with massive
propagating ({\it not} auxiliary)
singlet mesons $N^r_s$.  We will give the mesons a slightly
unusual mass term, and couple them to the other fields; the
superpotential is
\be\label{MMopB}
W=\lam N_r^s\fQn{r}\aQn{s}
+{1\over 2}m_0\left[ N_r^s N_s^r-\frac{1}{\nc}N_r^r N_s^s\right]
\ee
We have seen this superpotential in \eqr{MMop} during our study of
the \djn model, which contains auxiliary fields $N^r_s$.
The SQCMD model has an anomaly-free
$U(2\nc)\times U(1)_R$ flavor symmetry.   Note that in the
limits $m_0=0$ and $m_0=\infty$ the flavor symmetry is enhanced
to $SU(2\nc)\times SU(2\nc)\times U(1)\times U(1)_R$.

There is a unique flavor-independent gauge-anomaly-free $R$ charge
with $R(Q)=R(\ot Q)=\hlf$ and $R(N)=1$.  This charge determines the
dimensions of chiral operators at any interacting fixed point.
The theory with $m_0=\infty$ is N=1 SQCD with $\nf=2\nc$;
according to \cite{NAD} it flows to an interacting
superconformal fixed point.  The operator
\be\label{QQQQopB}
\left[(Q^r_\al \ot Q_s^\al)(Q^s_\bt \ot Q_r^\bt)
-\frac{1}{\nc} (Q^r_\al \ot Q_r^\al)(Q^s_\bt \ot Q_s^\bt)\right] \ ,
\ee
seen earlier in \eqr{QQQQop} during our study of the \djn model,
is exactly marginal at this fixed point; its \bata\ is $A_h=1+2\gm_Q$
while the gauge \bata\ is $A_g=-(\nc+2\nc\gm_Q)$.

Of course, the \djn model, in the
limit in which the N=2 gauge coupling is taken to zero and $m_\Sigma$
is taken to $\infty$, also becomes the N=1 SQCD theory studied in
\cite{NAD}; it flows to the same fixed point, and has the same marginal
operator in the infrared, as the SQCMD model with $m_0=\infty$.

The operator \eqr{QQQQopB} generates a complex curve of fixed points.  We
may flow to this curve in two ways.  One way is to take the \djn
model with finite N=2 gauge coupling $\tau$, as seen in \eqr{QQQQop}.
The other is to let the mass $m_0$ in SQCMD be finite but large;
integrating out the meson we generate the operator \eqr{QQQQopB}.
The fact that in the ultraviolet the meson is an auxiliary field
of dimension 2 in the \djn model and a propagating canonical field
of dimension 1 in SQCMD is unimportant;
by the time the two theories have reached the infrared, the dimension
of each meson has flowed to $\frac{3}{2}$ and the two theories are
indistinguishable.

Generally, we do not know what value of $h$ will be found at the
low-energy fixed point when flowing
from the \djn or SQCMD theories with given initial values
of $\tau$ or $m_0$.  There is one exception: when $\tau\rarr i\infty$ or
$m_0=\infty$, then $h=0$.  However, we also know, by holomorphy, that
continuous variations in $\tau$ or $m_0$ will generically lead to continuous
variations in $h$, so for sufficiently large values of ${\rm Im}\ \tau$ or
of $m_0$, the low-energy coupling $h$ will be small.

The structure of the fixed curve may be explored using SQCMD.  Our
key point is the following. From $m_0=\infty$
the mass of the meson may be continued to zero without the theory
leaving the fixed curve.  This follows from the fact that the
meson has $R=1$, so that its mass preserves the $R$ symmetry; thus
the theory has the same $R$ symmetry in the infrared independent
of whether $m_0$ is zero, finite, or infinite.  This is crucial.
In general, a mass term is a relevant perturbation on a fixed point
and causes the theory to flow to a new one.  Here it is marginal,
and the infrared theory remains superconformal for any $m_0$.
There are two distinct limits.  When $m_0$ is much larger than the
dynamical scale $\Lam$ of the theory, it is appropriate to integrate
out the meson and think of the theory as SQCD with the superpotential
\eqr{QQQQopB}.  When $m_0\ll\Lam$, the meson mass is negligible
in the ultraviolet, while in the infrared it becomes a dimensionless
coupling constant --- it undergoes dimensional anti-transmutation.
In this case the meson should remain in the theory and the superpotential
\eqr{MMopB} is appropriate.  But there is no dividing line between these
descriptions and no reason whatever that they should not go smoothly
into one another.  We therefore conclude that the mass $m_0$
parameterizes a complex curve of fixed points which connects the theory
with $m_0=0$ to that with $m_0=\infty$.
Note that we have not demanded that this complex curve (of real
dimension two) be everywhere
non-singular.  We will require only that there is a smooth path
(of real dimension one) connecting $m_0=\infty$ ($h=0$) to $m_0=0$.

We have noted that the theories at infinite and zero $m_0$ have enhanced
flavor symmetries.  Could they be the same point?  The answer is no.
The gauge invariant operators of the two theories do not match, as
can be seen from  consideration of \cite{NAD}.  The baryons
\be
B^{r_1\cdots r_N} \equiv
\eps_{\al_1\cdots\al_N}Q^{r_1\al_1}\cdots Q^{r_N\al_N} \ ,
\ee
with unit baryon number,
exist all along the fixed curve and are the same operators at
both endpoints, as are the antibaryons.
At $m_0=\infty$, the gauge invariant mesons are
$Q^{r\al}\ot Q_{s\al}$ which have zero baryon number and transform as
$({\bf 2N_c},{\bf\br{2N}_c})$ under the
$SU(2\nc)\times SU(2\nc)\times U(1)$ flavor symmetry.
For finite $m_0$ the flavor symmetry is broken by the superpotential
to $U(2\nc)$;
the operators $Q^r\ot Q_s$ and $N^s_r$, both of which transform as a
$({\bf 2N_c}\otimes {\bf\br{2N}_c})$, are mixed by the equations of motion.
At $m_0=0$, the larger flavor symmetry is again present;
$\fQn{r}\aQn{s}$ is a redundant operator, and the only other gauge
and baryon singlets are $N^s_r$, which transform as a
$({\bf\br{2N}_c},{\bf 2N_c})$ under the flavor symmetry.  No
manipulation of the flavor symmetries can simultaneously
make the baryons and mesons of the two theories match.

We have already seen that the limit $\tau\rarr i\infty$ (weak coupling)
in the \djn model leads to the same fixed point as does SQCMD with
$m_0=\infty$.  What point does $m_0=0$ correspond to?  It must be a
special point in the \djn model, since it has an enhanced flavor
symmetry.  As the adjoint field is flavor-blind, we expect that such
a point must derive from a special point in the N=2 theory.
Only the free theory $\tau\rarr i\infty$ is known to have an
enhanced $SU(2\nc)\times SU(2\nc)\times U(1)$ flavor symmetry; at generic
values of $\tau$ the N=2 model has merely a $U(2\nc)$ symmetry.
A reasonable guess is that the theory has enhanced symmetry at infinite
coupling ($\tau\rarr 0$), a point which should be related by S-duality to
a free theory (of magnetic matter) which would have
an enhanced flavor symmetry due to the absence of interactions.

We end this section by summarizing our conclusions, depicted in
Fig.~\ref{fig:SQCMDfig}.  We have found that
the fixed curve of the \djn model is the same as that generated by
the meson mass operator \eqr{MMopB} in SQCMD.  There exists
a path along which one may go smoothly from
$m_0=\infty$ to $m_0=0$; the endpoints of this path are inequivalent,
though they share the same enhanced flavor symmetry.  These points
may also be reached from the \djn model, the former in the limit
of weak coupling, the latter in the limit of strong coupling.

\subsection{N=2 duality and its effects}

We now consider the action of S-duality on these theories.  Much of
what follows relies on several assumptions about its effects in N=2 theories
for $SU(\nc >2)$, since the duality of these theories has not yet
been fully described.  These assumptions are in part tailored to
assure a relation between N=1 and N=2 duality, so we are not
proving anything here.  Still, our assumptions are consistent
with the known case \cite{NSEW} and with other
properties of these models, and we believe that our assumptions are
at least in part correct.

The main assumption is that these finite models have duality under
$\tau\leftrightarrow -1/\tau$.  Let us consider what form this
$\ZZ_2$ symmetry could take.  It is clear the dual theory must also be
a finite N=2 theory.  However, of all these
theories, only the $SU(\nc)$ model has a $U(2\nc)$ flavor symmetry for
arbitrary $\nc$. [The flavor symmetries of finite $SO(\nc)$ and
$Sp(2\nc)$ models with vector multiplets are $Sp(2\nc-4)$
and $SO(4\nc+4)$.]  We therefore infer
that the $SU(\nc)$ theory must be transformed into another theory with
the same gauge and matter content.  This is of course true in the known
case of $SU(2)$\cite{NSEW}.  It is also encouraging that the
semi-classical  monopoles of the broken $SU(\nc)$ theory include states
in the ${\bf 2\nc}$ and ${\bf\br{2N}_c}$ of flavor.  There are also
other monopoles in larger flavor representations; these states cannot
be present in the unbroken N=2 theory as they would carry color and would
give the theory a positive $\bt$-function.

In the $SU(2)$ case \cite{NSEW} the massless particles at the
origin include the eight massless monopoles in the spinor of $SO(8)$.
We will guess that the generalization of this statement is that
there are monopole-like states $q_r$ and $\ot q^s$ in the ${\bf 2\nc}$ and
${\bf\overline{2N}_c}$ of flavor which are massless in the
unbroken theory.  The need for a vanishing
one-loop $\bt$-function forces us to put these fields into
representations of index one; we will take $q_r$
in the ${\bf \nc}$ of the dual $SU(\nc)$ and $\ot q^s$ in the conjugate
representation.

We must next consider the operator mapping in the N=2 model under
S-duality.  We begin at an arbitrary coupling $\tau$, where
the flavor symmetry is $U(2\nc)$.  As in the \djn model
the gauge invariant operators include meson and baryon operators
built from the invariant tensors of $SU(\nc)$.  These same operators
must reappear in the dual description of the theory.

Of course, the simplest way for this
to occur would be for the fields $Q^r,\ot Q_s$ to be mapped to the fields
$\ot q^r,q_s$, and for each operator built out of $Q^r$ fields
to be mapped to the same operator built out of $\ot q^r$ fields.  In this
case the duality would simply map $\tau\rarr -1/\tau$
and leave the flavor representations unchanged.  Also, $\ot q^r$ would
have positive baryon number.
However, this hypothesis would lead to an
inconsistency with a property of SQCMD discussed in the previous section.
If there were no transformation of the flavor structure under N=2 duality,
then the fixed line of the N=2 model would be dual to itself,
with the theory at $\tau$ identical to the theory at $-1/\tau$.  In
this case the fixed curve of the \djn model would be dual to itself;
but we have already argued that the theories derived from $\tau=0$
and $\tau=i\infty$ were inequivalent using the properties of SQCMD.
We therefore discard the hypothesis that the flavor structure is
unchanged under duality.

Since the mapping $\tau\rarr -1/\tau$ is a $\ZZ_2$ transformation,
any other hypothesis must involve a $\ZZ_2$ transformation of the
flavor representations.  In this case the N=2 theory is mapped
under $\tau\rarr -1/\tau$ to an inequivalent but similar N=2* theory,
with a fixed line that contains the same physics except conjugated
by a $\ZZ_2$.  A natural candidate involves a $\ZZ_2$ automorphism of
$U(2\nc)$ --- charge conjugation ---  which maps $SU(\nc)$
invariants as follows.
\be\label{opmap}\begin{array}{rcl}
 Q^{r_1}\cdots Q^{r_\nc}& \leftrightarrow &
\eps^{r_1\cdots r_\nc s_1 \cdots s_\nc}  q_{s_1} \cdots  q_{s_\nc}\\ \\
\ot Q_{s_1}\cdots \ot Q_{s_\nc}& \leftrightarrow &
\eps_{s_1\cdots s_\nc r_1 \cdots r_\nc} \ot q^{r_1} \cdots \ot
q^{r_\nc}\\ \\
Q^r\ot Q_s -\frac{1}{2\nc}\delta^r_s Q^u\ot Q_u&
\leftrightarrow &
 -\left[ q_s \ot q^r - \frac{1}{2\nc}\delta^r_s q_u\ot q^u\right]\\ \\
Q^r\ot Q_r& \leftrightarrow &  + q_r \ot q^r
\end{array}
\ee
Under this transformation the $q_r$ have positive baryon number.
Note that the operator map cannot include the adjoint superfield
$\Sigma$ by symmetries and dimensional analysis.

Furthermore, the operator $\Sigma^2$ must map (again by symmetries
and dimensional analysis) to itself
under duality, so a $\Sigma$ mass term in the N=2 theory will map to
one in the N=2* theory.  It follows that the \djn model will be mapped to an
inequivalent \djns model under S-duality.
Of course, the N=2
supersymmetry is broken when the adjoint field is massive, so
we expect N=2 duality to be broken also.  However, it is reasonable
to expect, since N=1 supersymmetry is still preserved,
that these violations of duality will occur in the
K\"ahler potential, and that the duality of the superpotential and of
the chiral operators will survive into the \djn and
\djns models.  This means that their infrared fixed points
should be dual, and that the operator map \eqr{opmap} should  be
appropriate for the \djn and \djns theories.

Although we do not know the exact map between the couplings
$h$ and $h^*$ (since both coupling
constants can be non-perturbatively renormalized), we do know, from
the enhanced flavor symmetries, that
the point $h=0$ on the \djn fixed curve
must be mapped to a special point $h^*\neq 0$ on the \djns fixed curve.
A natural guess is that the special point is at $h^*=\infty$.
Similarly the point $h^*=0$ would be mapped under S-duality to $h=\infty$.
Thus, qualitatively, the small and large coupling regions exchange
places.  In terms of the auxiliary meson introduced in \eqr{MMop},
the large mass region of the \djn model is dual to the small mass region
of the \djns model, with the points $m_0=0,\infty$ exchanged with
$m_0^*=\infty, 0$.  This qualitative picture is shown in
Fig.~\ref{fig:NisIIdual}.

\subsection{SQCMD and N=1 duality}

By now the reader can clearly see where we are heading.  The map
between the \djn and \djns fixed curves induces a map between the
SQCMD theory studied earlier and a dual theory SQCMD*.  The latter has
fields $q_r,\ot q^s$ in the fundamental and antifundamental
representations of $SU(\nc)$ and singlet mesons $M^r_s$ of
mass $m_0^*$.  The map
between the theories is not applicable in the ultraviolet;
it is merely a map between their infrared fixed points.

Now consider a point on the fixed curve of SQCMD
where $m_0$ is large.  This point is mapped
to a point with small $m_0^*$ on the SQCMD* fixed curve.  For large $m_0$
it is appropriate to integrate out the meson field, leaving
the superpotential \eqr{QQQQopB} with a small coupling $h$.
The gauge invariant chiral operators of the theory are
\be
 Q^{r_1}\cdots Q^{r_\nc}, \ \
 \ot Q_{s_1}\cdots \ot Q_{s_\nc}, \ \
Q^r\ot Q_s  \ .
\ee
The dual theory has superpotential
\be
W_*=\lam^* M^r_s q_r\tilde
q^s+{1\over2}m_0^*\left[M^s_rM^r_s-{1\over\nc}M^r_rM^s_s\right]
\ee
and contains the operators
\be\label{opsB}
q_{r_1} \cdots  q_{r_\nc}, \ \
 \ot q^{s_1} \cdots \ot q^{s_\nc}, \ \
 q_r \ot q^s , \ \
M^r_s \ .
\ee
Note the mixing between the singlets $M^r_s$ and $q_r\ot q^s$ through
the infrared equations of motion:
\be\label{qMeq}
q_r\ot q^s = -\frac{m_0^*}{\lam^*}
\left[M^s_r-\frac{1}{\nc}\delta^s_r M^u_u\right]
\ee
The operator map \eqr{opmap} takes $Q^r\ot Q_s$ to
$q_s\ot q^r$ and thus to $M^r_s$.  The massless particles in the
two theories are $Q^r, \ot Q_s$ and $q_s,\ot q^r$; one may verify
that the 't Hooft anomaly matching conditions
are trivially satisfied.

What happens in the limit $m_0\rarr \infty$?  We argued earlier that
this theory should be dual to $m_0^*=0$ simply from the enhanced
flavor symmetries.  The superpotential of the SQCMD theory
is zero while that of the SQCMD* theory is $W=\lam^* M^r_s q_s\ot q^r$.
The operator $q_r\ot q^s$ is now redundant
and disappears from the spectrum.  The operator map \eqr{opmap}
becomes, using \eqr{qMeq},
\be\label{opmapC}\begin{array}{rcl}
 Q^{r_1}\cdots Q^{r_\nc}& \leftrightarrow &
\eps^{r_1\cdots r_\nc s_1 \cdots s_\nc}  q_{s_1} \cdots  q_{s_\nc}\\ \\
\ot Q_{s_1}\cdots \ot Q_{s_\nc}& \leftrightarrow &
\eps_{s_1\cdots s_\nc r_1 \cdots r_\nc} \ot q^{r_1} \cdots \ot
q^{r_\nc}\\ \\
Q^r\ot Q_s & \leftrightarrow &  M^r_s
\end{array}
\ee
The massless particles of the theories are $Q^r, \ot Q_s$
in SQCMD and $q^r,\ot q_s, M^r_s$ in SQCMD*.  One may confirm \cite{NAD}
that these particles satisfy the 't Hooft anomaly matching conditions
under the now enhanced
$SU(2\nc)_L\times SU(2\nc)_R\times U(1)_B\times U(1)_R$ global symmetry.

The description of the previous paragraph coincides precisely with
the N=1 duality map
introduced by Seiberg \cite{NAD} for the case $N_f=2\nc$.  A diagram
illustrating our arguments appears in Fig.~\ref{fig:diagram}.

\subsection{Other aspects of N=1 duality}

We have identified a possible source for N=1 duality in the
$SU(\nc)$ theory studied by Seiberg \cite{NAD} with $\nf=2\nc$.
To a certain degree this is sufficient for $\nf\neq 2\nc$ as well.
One may move from the known theory to any other by perturbing it
by relevant operators, and, knowing how operators
are mapped under N=1 duality at the initial point, we know how any
given perturbation acts both in the original and in the dual theory.
However, the full duality also involves non-perturbative effects
\cite{NAD} which our methods have not identified.  What we may
say with confidence is that the mere
uniqueness of the anomaly-free $R$ charge following perturbation by
a relevant operator automatically guarantees the correct
mapping of operators at a new fixed point and the matching of
global anomalies found in \cite{NAD}.  This can be seen by
considering infinitesimal mass and symmetry breaking
perturbations, without integrating out any massive fields.
The new flavor symmetries, and therefore the new
flavor anomalies, are linear combinations
of the old ones; the linear combinations in
the original theory are the same as those in the dual theory,
so the new anomalies match.  Those fields which
are massive when the perturbations are finite have cancelling
anomalies, of course, so when they are integrated out the anomalies
still match.  Similar arguments can be used for the operators.
These arguments apply when both the original and dual theory
flow to an interacting fixed point, which includes the cases
$\frac{3}{2}\nc<\nf<3\nc$.

For larger $\nf$ the one-loop
$\bt$-function is positive and the infrared fixed point must
be free, while for smaller $\nf$ there is a breakdown of
unitarity \cite{NAD} which implies that the description in terms of
the original fields must fail. This and other subtleties
(strong coupling, non-perturbative
effects, confinement) are important for understanding
the properties of and sometimes even the existence of
certain fixed points.  Insight into
these issues might be gained by studying N=2 theories more thoroughly.

\subsection{Commentary}

We have presented a mechanism by which the S-duality of finite N=2 models
could survive perturbation by an N=2 breaking term and could
be transferred to theories which preserve only N=1 supersymmetry.
We did so by giving a mass to the adjoint chiral superfield
and showing that the resulting \djn model had a marginal operator
whose associated fixed curve was identical to that of another
model, SQCMD.  We noted that the endpoints of the curve ($m_0=\infty,0$) had
enhanced flavor symmetries, and we used this fact to argue that
the weak and strong coupling limits of the N=2 theory flowed to
these two endpoints.  By showing that the two endpoints were distinct
theories, we found ourselves forced to conclude that S-duality
could not map the N=2 theory onto itself.  However, the large flavor
symmetries  of the N=2 theory
strongly suggested that S-duality maps it to an N=2 theory with
the same color and flavor representations.  Since the mapping
$\tau\rarr -1/\tau$ is a
$\ZZ_2$ transformation, we were left with only one option; the
duality transformation must act as a $\ZZ_2$ on the flavor space.
Fortunately, there was a natural  $\ZZ_2$ of the flavor $U(2\nc)$ ---
charge conjugation --- which acted on operators in a simple way.
When we brought
our conjecture for the S-duality transformation of the N=2 theory into
the \djn model, and thereby into the SQCMD model, we found that
the fixed curves of the SQCMD model and the dual SQCMD* model
were mapped to one another, with large and small meson mass regions
exchanging places.  In particular the zero mass and infinite mass
endpoints were mapped to one another.  Analysis of the operator
identifications, the massless fields, and the superpotentials showed
that the mapping of the endpoints was precisely the N=1 duality mapping
found by Seiberg \cite{NAD} in the particular case $\nf=2\nc$.

There are a number of gaps in our reasoning.  We have assumed that
the structure of the fixed curves is fairly straightforward and
that there is no obstruction to traveling from $m_0=\infty$ to
$m_0=0$ in the SQCMD model.  We did not prove that
the strong coupling limit $\tau=0$ of the N=2 model flows to the $m_0=0$
endpoint of the SQCMD model, which is critical for the argument.
We also did not prove, much less derive, that the N=2 model is
dual to itself up to a flavor transformation.
Our choice for the $\ZZ_2$ flavor transformation appears to be unique,
but perhaps there is a more subtle option.
We have also neglected other aspects of N=2
duality, such as the existence of dyons.  Insight into this
and other related issues is to be found in \cite{NSKItwo}.
We are optimistic that a
substantial part of our mechanism is correct, given the consistency
of our picture with the various known examples \cite{NAD,NSKItwo,NSEW}.

Assuming that we have correctly identified the physical mechanism
underlying N=1 duality, there are a number of comments to be made.
First, the picture apparently extends to $SO(\nc)$ and $Sp(2\nc)$ models
with vector multiplets; certainly most of the arguments go through.
In the case of $Sp(2\nc)$ there are no new issues.  However, for
$SO(\nc)$, the flavor symmetry is $Sp(2\nc-4)$, and a $\ZZ_2$
reflection in this space would not transform baryons ($\nc$-index
antisymmetric tensors) into baryons in the dual theory.  A more
intricate operator mapping than Eq.~\eqr{opmap} will therefore be
necessary.  This is presumably related to the complexity of
N=1 duality for $SO(\nc)$ as compared with $SU(\nc)$.\cite{NAD}

It is natural to try to generalize our
mechanism to finite  N=2 theories with more complicated representations;
however, there are some obstacles.  When
superfields are integrated out of other N=2 models,
the Fierz transformation used in \eqr{QQQQop} leads to a sum of a terms,
not all of which can be written as a product of bilinear invariants.
The introduction of a singlet auxiliary meson superfield then
leaves other non-renormalizable operators in the superpotential whose
coupling constant becomes large in the limit that the meson mass
becomes small.
Another problem is that the duality of N=2 models is even less well
understood for other representations.  For example, the gauge group of
the dual theory is not necessarily the same as that of the original;
indeed this is expected to occur in N=4 models.

If it is true that all N=1 duality stems exclusively from N=2 duality,
the situation appears unfortunate for chiral models, which cannot be
derived from N=2 theories.  However, the finiteness of N=4
and N=2 models generalized to both vector-like and chiral N=1
models.  Also, the $E_6$ model of
Sec.~\ref{subsec:Esix} bore a close resemblance to the $SU(3)$
model of Sec.~\ref{subsec:SUthree}, which had a duality transformation
(to the model of Sec.~\ref{subsec:SUsix}) on its
fixed curve.  This leads us to wonder whether even chiral models with
fixed curves might have some form of weak-strong coupling duality.

One may also speculate as to whether the mechanism we have identified
can be generalized in some way to bring N=1 duality down to
non-supersymmetric theories.  One might consider integrating out
the gluinos, for example.  We hope to return to this question in the
future.

\section{Conclusion}
\label{sec:conc}

We have shown that exactly marginal operators and extended manifolds
of fixed points are commonplace in N=1 supersymmetric gauge theory.
Our approach treats the previously known finite gauge theories and
the specific case of $SO(4)$ studied in
\cite{NAD,NSKItwo} on an equal footing.  Many new
examples were presented.  Our exploration of models has not been by
any means exhaustive, but
we have found a wide range of interesting theories which can be expected
after further study to reveal a variety of new phenomena.
We saw that manifolds of fixed points
may be weakly or strongly coupled; they may be generated by
renormalizable or non-renormalizable operators; theories containing
them may be chiral or vector-like.  Our examples
have included finite models, models with curves of fixed points which
are everywhere weakly coupled but nowhere free, models with
strongly coupled manifolds of fixed points, and models which
have fixed curves which run to infinite gauge coupling.
The limitations of our
methods are still unclear, but the consistency of the picture we
advocate here is strong evidence that they hold in many interesting cases.

The issue of duality has come up in several contexts.
We have pointed out several finite N=1 models that may have S-duality
of some form.  The N=1 duality of Seiberg \cite{NAD,NSKItwo} has
been essential in giving us confidence that manifolds of fixed points
are really present in non-renormalizable models.
We have also identified many models in which
the fixed curves of finite models flow to or from fixed
curves of theories that are perturbatively non-renormalizable;
those which stem from finite N=2 models may well have a form
of N=1 duality.

Of these, the models studied in \cite{NAD} with gauge
group $SU(\nc)$ and $\nf=2\nc$ flavors are of great interest.  It was
suggested in \cite{NAD} that the N=1 duality studied therein might be
related to duality of the finite N=2 model with  $\nf=2\nc$.
We have suggested a physical mechanism by which this relationship could
be realized.  The electric and magnetic descriptions
of the N=2 fixed curve flow under an N=1 breaking perturbation
to dual descriptions of an N=1 fixed curve.  With
a few reasonable assumptions, we have demonstrated that
one point of this curve is described by the electric and magnetic
theories of \cite{NAD} with $\nf=2\nc$.
This certainly does not constitute a derivation of N=1 duality
from N=2 duality; our arguments are speculative.  Still, their
consistency with Refs.~\cite{NAD} and \cite{NSEW} makes us optimistic that
they will prove, at least in part, to be correct.

The exploration of the space of four dimensional conformal field
theories has just begun.  We hope that, as in two dimensions, the
study of marginal operators will open many new paths for investigation.

\bigskip

It is a pleasure to thank our colleagues M. Douglas, D. Friedan,
K. Intriligator, N. Seiberg  and S. Shenker for their advice and
encouragement.  We also had useful and enjoyable discussions with
M. Dine, M. Peskin, E. Silverstein and C. Vafa.

\figure{\label{fig:Nisfour}
The renormalization group flow near the N=4 fixed line, shown
schematically, in the plane
of the gauge coupling $g$ and the Yukawa coupling $h$.  Arrows indicate
flow toward the infrared.}

\figure{\label{fig:GcrossG}
The renormalization group flow near the fixed curve of the model
of section Sec.~\ref{subsec:SptimesSp}, shown schematically, in the plane
of the two gauge couplings.  Arrows indicate flow toward the infrared.}

\figure{\label{fig:SUiv}
The renormalization group flow near the fixed curve associated
with the superpotential \eqr{sufourB}, shown
schematically, in the plane of the gauge coupling $g$ and the Yukawa
coupling $h$.  Arrows indicate flow toward the infrared.}

\figure{\label{fig:SUvi}
The renormalization group flow near the fixed curve of the model
of section Sec.~\ref{subsec:SUsix}, shown
schematically, in the plane
of the gauge coupling $g$ and the Yukawa coupling $h$.  Arrows indicate
flow toward the infrared.}

\figure{\label{fig:SQCMDfig}
The SQCMD theory flows to the same fixed curve as the \djn model (a
finite N=2 model broken by a mass term $m_\Sigma \Sigma^2$ for the
adjoint chiral superfield.)   Along the fixed curve the flavor
symmetry is $U(\nc)$, except at the endpoints where it is
$SU(\nc)\times SU(\nc)\times U(1)$. The theory at $h=0$
$(m_0=\infty,\tau\rarr i\infty)$ is
different from that at $h=\infty$ $(m_0=0=\tau)$.}

\figure{\label{fig:NisIIdual}
The duality of N=2 maps the electric theory with $\tau$ to the
magnetic N=2* theory with $\tau^*=-1/\tau$.  The duality is carried
down to the infrared fixed curves of the \djn and \djns models.  Note
that the endpoints of the fixed curves are mapped to one another.}

\figure{\label{fig:diagram}
The duality of the \djn and \djns models translates into an infrared duality
for the SQCMD and SQCMD* models.  The theory at $m_0=\infty$ is equivalent
under duality to the theory at $m^*_0=0$; these limits of SQCMD and
SQCMD* were first identified as dual by Seiberg \cite{NAD}.}

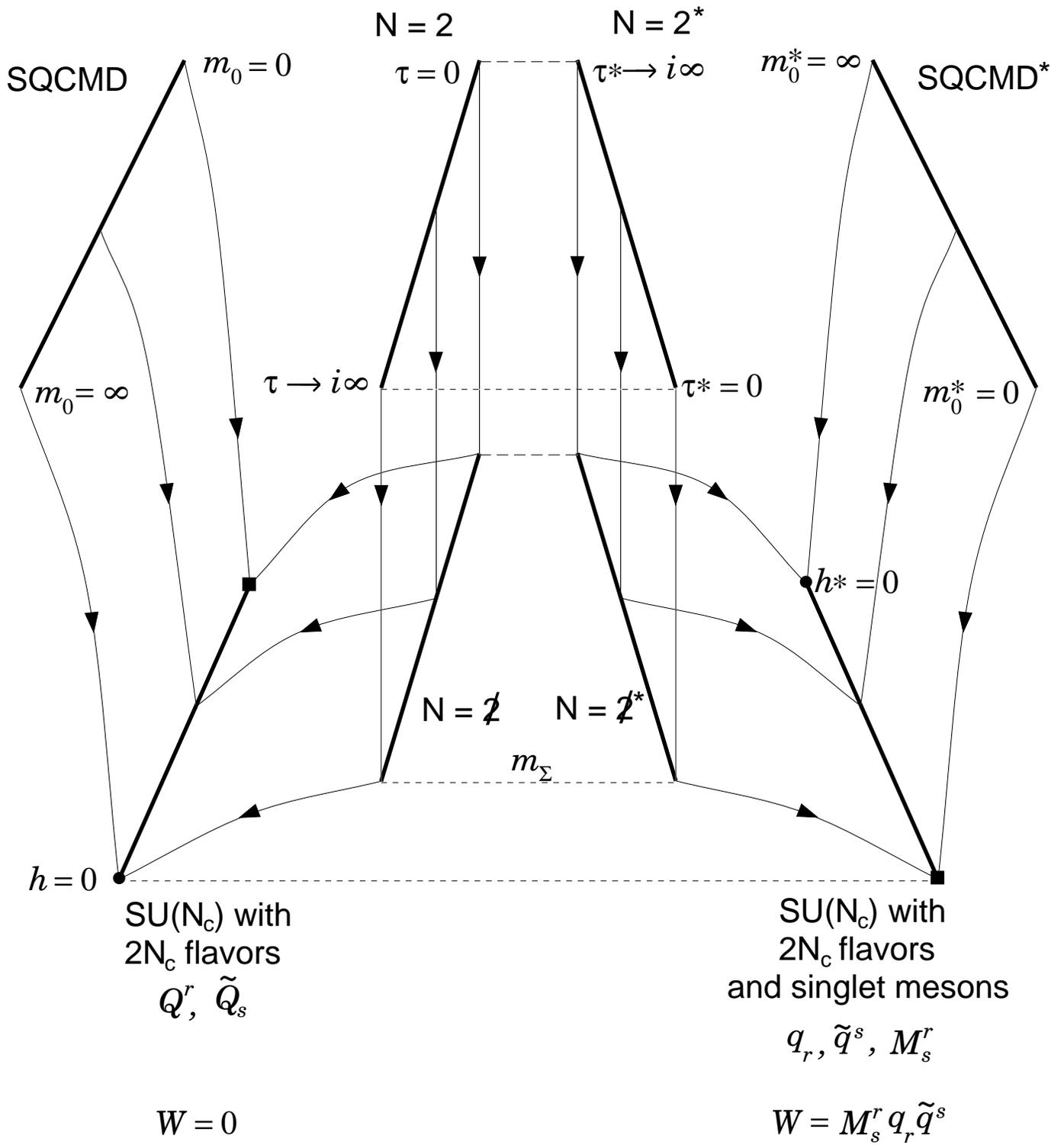
\end{document}